\documentclass[aps, prd, 10pt, twocolumn, superscriptaddress,noshowpacs, preprintnumbers, longbibliography, bibnotes,hyperref,floatfix]{revtex4-1}
\usepackage[colorlinks=true,breaklinks=true]{hyperref}
\usepackage{color}	    
\hypersetup{allcolors=[rgb]{0.0 0.0 1.0},linkcolor=[rgb]{0.75 0.05 0.05}}
\usepackage[dvipsnames]{xcolor}

\usepackage{listings}
\usepackage{mathtools}
\usepackage{amsmath}
\usepackage{amsfonts}
\usepackage{amssymb}
\usepackage{bbold}
\usepackage{epsfig}
\usepackage{graphicx}
\usepackage{bm}
\usepackage{array}
\usepackage{hyperref}
\usepackage{listings}
\usepackage{color}
\usepackage{float}
\usepackage[normalem]{ulem}

\newcommand{\GF}{G_{\rm F}}

\newcommand{\sD}{{\sf D}}
\newcommand{\sS}{{\sf S}}

\newcommand{\vR}{\boldsymbol{R}}
\newcommand{\vJ}{\boldsymbol{J}}
\newcommand{\vS}{\boldsymbol{S}}

\newcommand{\vD}{\boldsymbol{D}}

\newcommand{\vP}{\boldsymbol{P}}
\newcommand{\vPb}{\boldsymbol{\bar{P}}}

\newcommand{\cmin}{\cos{\vartheta}_{\mathrm{min}}}

\newcommand{\exclude}[1]{}

\usepackage{tikz,xcolor,hyperref}

\definecolor{lime}{HTML}{A6CE39}
\DeclareRobustCommand{\orcidicon}{\hspace{-1mm}
	\begin{tikzpicture}
	\draw[lime, fill=lime] (0,0) 
	circle [radius=0.16] 
	node[white] {{\fontfamily{qag}\selectfont \tiny \,ID}};
	\draw[white, fill=white] (-0.0525,0.095) 
	circle [radius=0.007];
	\end{tikzpicture}
	\hspace{-3mm}
}

\foreach \x in {A, ..., Z}{\expandafter\xdef\csname orcid\x\endcsname{\noexpand\href{https://orcid.org/\csname orcidauthor\x\endcsname}
			{\noexpand\orcidicon}}
}

\begin{document}

\preprint{MPP-2022-120}

\title{Neutrino Fast Flavor Pendulum. Part 2:~Collisional Damping}

\author{Ian Padilla-Gay\orcidA{}}
\affiliation{Niels Bohr International Academy \& DARK, Niels Bohr Institute,\\University of Copenhagen, Blegdamsvej 17, 2100 Copenhagen, Denmark}
\author{Irene Tamborra\orcidB{}}
\affiliation{Niels Bohr International Academy \& DARK, Niels Bohr Institute,\\University of Copenhagen, Blegdamsvej 17, 2100 Copenhagen, Denmark}
\author{Georg G. Raffelt\orcidC{}}
\affiliation{Max-Planck-Institut f\"ur Physik (Werner-Heisenberg-Institut),\\ F\"ohringer Ring 6, 80805, Munich, Germany}

\date{\today}

\begin{abstract}
In compact astrophysical objects, the neutrino density can be so high that neutrino-neutrino refraction can lead to fast flavor conversion of the kind $\nu_e \bar\nu_e \leftrightarrow \nu_x \bar\nu_x$  with $x=\mu,\tau$, depending on the neutrino angle distribution. Previously, we have shown that in a homogeneous, axisymmetric two-flavor system, these collective solutions evolve in analogy to a gyroscopic pendulum. In flavor space, its deviation from the weak-interaction direction is quantified by a variable $\cos\vartheta$ that moves between $+1$ and $\cos\vartheta_{\rm min}$, the latter following from a linear mode analysis. As a next step, we include collisional damping of flavor coherence, assuming a common damping rate $\Gamma$ for all modes. Empirically we find that the damped pendular motion reaches an asymptotic level of pair conversion $f=A+(1-A)\cos\vartheta_{\rm min}$ (numerically $A\simeq 0.370$) that does not depend on details of the angular distribution (except for fixing $\cos\vartheta_{\rm min}$), the initial seed, nor $\Gamma$. On the other hand, even a small asymmetry between the neutrino and antineutrino damping rates strongly changes this picture and can even enable flavor instabilities in  otherwise stable systems.
\end{abstract}

\maketitle

\section{Introduction}
\label{sec:intro}

The flavor evolution of neutrinos and antineutrinos in neutrino-dense environments is nonlinear because of the coherent forward scattering of neutrinos among each other~\cite{Pantaleone:1992eq}. This phenomenon is commonly referred to as collective neutrino oscillation and can span over different time and length scales~\cite{Duan:2010bg, Mirizzi:2015eza, Tamborra:2020cul, Capozzi:2022slf, Richers:2022zug}. The neutrino and antineutrino medium supports different types of collective flavor evolution, depending on the flavor-dependent energy and angle distributions. A normal-mode analysis in the linear regime reveals the conditions for the existence of unstable (``tachyonic'') modes that are characterized by an eigenfrequency with  nonvanishing imaginary part \cite{Banerjee:2011fj, Izaguirre:2016gsx, Airen:2018nvp, Morinaga:2021vmc, Dasgupta:2021gfs, Shalgar:2019kzy,Capozzi:2017gqd}.

In contrast to the early universe, compact astrophysical objects are characterized  by a large matter density that further affects flavor transformation: neutrino propagation states  nearly coincide with eigenstates of flavor instead of mass~\cite{Tamborra:2020cul,Richers:2022zug}. Accordingly, traditional numerical studies of neutrino transport in core-collapse supernovae (SNe) or binary neutron-star mergers ignore flavor correlations~\cite{Mezzacappa:2020oyq,Burrows:2020qrp,Janka:2012wk}. Neutrino transport of energy and lepton number in compact objects thus begs the question whether this approach remains viable in view of potentially strong collective flavor evolution that may occur despite, or in addition to, matter effects.

Much recent attention has revolved around ``fast flavor conversion'' (FFC) because it can operate on small length scales, driven by the neutrino-neutrino matter effect alone, driven by  $\mu=\sqrt{2}G_{\rm F}n_\nu$. The latter is much larger than the scale $\Delta m^2/2E$ that governs ``slow oscillations'' caused by the vacuum masses~\cite{Duan:2010bg}. Slow collective flavor transformation remains of interest in the regions where neutrinos stream away, e.g.~beyond the SN core. 

Pure FFC  (operating in the limit of vanishing vacuum frequency) acts on flavor lepton numbers and not on neutrinos and antineutrinos separately. In the mean-field approach, the flavor field is represented by $3\times3$ matrices $\varrho(t,{\bf r},{\bf p})$ as extensions of the usual occupation numbers, and $\bar\varrho(t,{\bf r},{\bf p})$ for antineutrinos. FFC, however, is self-consistently described by the lepton-number matrix $\sD(t,{\bf r},{\bf p})=\varrho(t,{\bf r},{\bf p})-\bar\varrho(t,{\bf r},{\bf p})$, whereas the solutions for $\varrho(t,{\bf r},{\bf p})$ and $\bar\varrho(t,{\bf r},{\bf p})$ follow from $\sD(t,{\bf r},{\bf p})$ that obeys a closed equation of motion (EOM)~\cite{Raffelt:2007yz,Padilla-Gay:2021haz,Johns:2019izj}.
 
The FFC effect in a two flavor system consisting of $\nu_e$ and $\nu_x=\nu_\mu$ or $\nu_\tau$ is driven by the angle distribution of the quantity $(f_{\nu_e}-f_{\bar\nu_e})-(f_{\nu_x}-f_{\bar\nu_x})$, often called the electron-lepton number (ELN) distribution, although this commonly adopted terminology is a bit misleading in that it is justified only if $f_{\nu_x}=f_{\bar\nu_x}$. (Notice, however, that in traditional three-species neutrino transport both $\nu_x$ and $\bar\nu_x$ have the same distribution function by construction.)
The initial FFC instability requires an ELN crossing, i.e., the angular ELN distribution must change sign for some direction~\cite{Izaguirre:2016gsx,Morinaga:2021vmc}. In the three-flavor context, the initial instability can occur in any of the two-flavor subsystems~\cite{Chakraborty:2019wxe}.

\begin{figure}
\centering
\includegraphics[width=1.0\columnwidth]{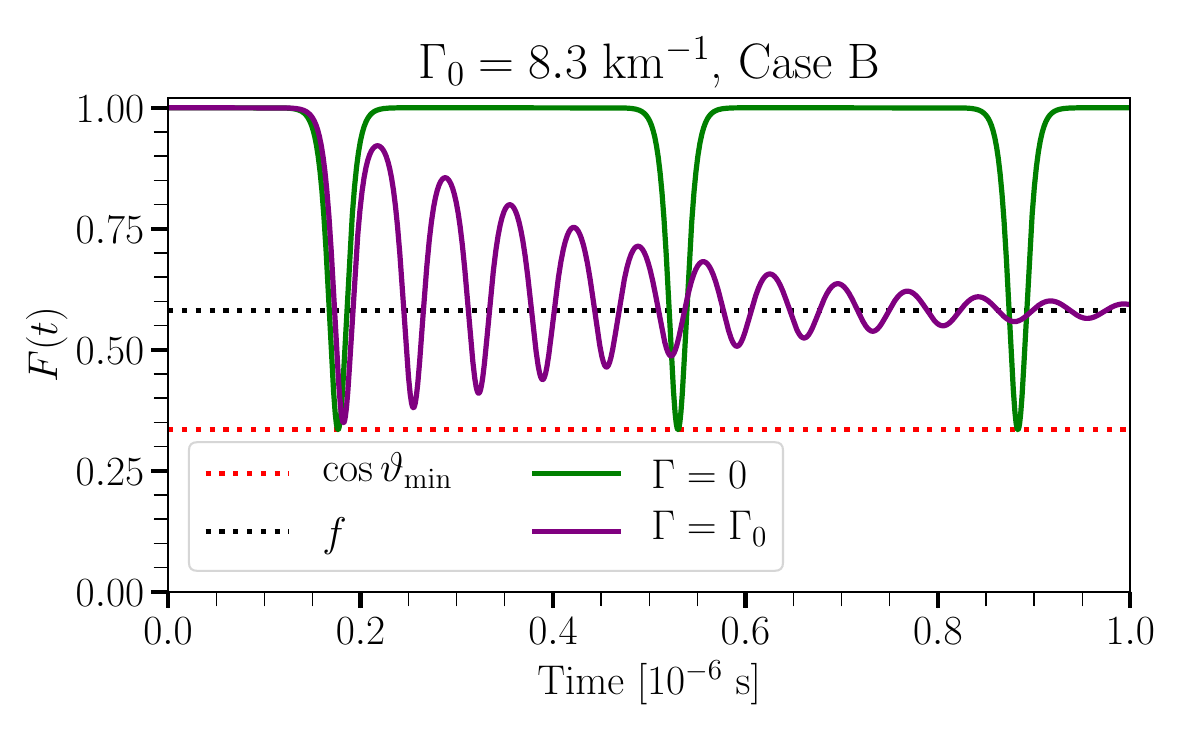}
\caption{Evolution for a typical fast-flavor pendulum (Case B of Paper~I) where we show $F(t)$ defined in Eq.~\eqref{eq:Ft} without damping (green) and with damping (purple). The length of the plateau phase depends on the smallness of the initial seed. The red dotted line shows the maximum excursion of the pendulum as in Eq.~\eqref{eq:cmin}, while the black dotted line marks the final steady state value $f$ as predicted by Eq.~\eqref{eq:asymptotic}.
}
\label{fig:dip}
\end{figure}

While the conditions for FFC instabilities can be found from a linear mode analysis, the possible astrophysical consequences of course depend on the nonlinear outcome that must rely on numerical simulations. One can still gain fresh insight from simple models that lend themselves to analytic discussion. We have recently explored such a case in Ref.~\cite{Padilla-Gay:2021haz} (Paper~I) in the form of an axially symmetric and homogeneous system, expanding on the formal analogy pointed out in Refs.~\cite{Hannestad:2006nj,Duan:2006an,Fogli:2007bk,Johns:2019izj}. We showed that the EOMs are formally equivalent to those of a gyroscopic pendulum. More specifically, the pendulating ``radius vector'' is the flavor polarization vector $\vD_1$ that describes the flavor lepton flux along the symmetry axis, whereas $\vD_0$, describing the flavor lepton density, is conserved and plays the role of ``gravity.'' The orientation of $\vD_1$ relative to the flavor-direction is quantified by the zenith angle $\vartheta$. Its motion is between the original unstable position $\vartheta=0$ (``upward'' position relative to ``gravity'') and later dips down to $\vartheta_{\rm min}$. A typical example is shown in Fig.~\ref{fig:dip}, corresponding to Case~B of Paper~I and also used later in this work.

The bottom turning point $\vartheta_{\rm min}$ of a gyroscopic pendulum is determined by its spin. One key finding of Paper~I is that $\vartheta_{\rm min}$ is given by the complex eigenfrequency of the linearized solution. For an unstable collective mode, it has the complex form $\omega_{\rm P}+i\gamma$ where $\omega_{\rm P}$ is the initial frequency of precession and $\gamma$ the exponential growth rate. The depth of the collective pair conversion~is
\begin{equation}\label{eq:cmin}
    \cos\vartheta_{\rm min}=\frac{\omega_{\rm P}^2-\gamma^2}{\omega_{\rm P}^2+\gamma^2}
    \ .
\end{equation}
This quantity therefore is found from the initial ELN distribution by performing a linear stability analysis, but without solving the nonlinear EOMs.

However, the physics of fast pairwise conversion remains full of unknowns. The short relevant length scales initially motivated studies that entirely ignored the effect of nonforward collisions. However, recent work has shown that collisions of neutrinos with the matter background can significantly affect FFC~\cite{Capozzi:2018clo, Johns:2021qby, Shalgar:2022rjj, Shalgar:2022lvv, Hansen:2022xza, Martin:2021xyl, Shalgar:2020wcx, Kato:2022vsu, Sasaki:2021zld}. In addition, the ELN angular distribution can be modified by neutrino advection dynamically~\cite{Shalgar:2022rjj, Shalgar:2022lvv, Shalgar:2019qwg, Richers:2021xtf, Nagakura:2022kic,Kato:2022vsu}.

We now venture to include the effect of collisions in the form of damping of flavor coherence, i.e., the off-diagonal elements of the flavor density matrices. The simplest case is that of a common damping rate $\Gamma$ for all modes. For a typical example, the damped pendular motion is shown in Fig.~\ref{fig:dip} (purple line), approaching an asymptotic value where the length of $\vD_1$ has shrunk and the system no longer moves. For convenience, we define the ELN flux factor
\begin{equation}\label{eq:Ft}
    F(t)=\frac{D_1^z(t)}{D_1^z(t_0)} \ ,
\end{equation}
whose asymptotic value as $t\rightarrow \infty$ we denote as
\begin{equation}\label{eq:F}
    f = \lim_{t \to \infty} F(t) \ .
\end{equation}
Notice that $F(t)=\cos\vartheta(t)$  in the absence of damping, whereas  the length of $\vD_1$ shrinks with damping, and the angle of $\vD_1$ with respect to  the $z$-direction is no longer the only relevant parameter.
We anticipate here our empirical finding of a universal connection
\begin{equation}\label{eq:asymptotic}
   f = A+(1-A)\cos\vartheta_{\rm min}\ ,
\end{equation}
where numerically $A\simeq0.370$. This result depends on initial conditions only through $\cos\vartheta_{\rm min}$, but not on $\Gamma$ or the initial seed.

This work is organized as follows. In Sec.~\ref{sec:eoms}, we present the EOMs of (anti)neutrinos and their multipole decomposition and introduce our specific ELN configurations that are identical to the ones adopted in Paper~I. Section~\ref{sec:impact_damping} focuses on the role of collisional damping, assuming a common rate $\Gamma$ for all modes. We consider the linear and nonlinear regime of flavor evolution and offer analytical estimates on how to compute the final flavor outcome based on initial ELN distributions. In Sec.~\ref{sec:diff_damping}, we consider different damping rates for neutrinos and antineutrinos and quantify the departure from the case of equal damping rates for particles and antiparticles. In Sec.~\ref{sec:dependence_ELN} we generalize our findings for a wide range of ELN distributions. Finally, closing remarks are reported in Sec.~\ref{sec:conclusions}. Appendix~\ref{appendix:lsa_continuous} outlines the normal mode analysis in the presence of collisional damping, while supplemental details on the numerical methods of our simulations are provided in Appendix~\ref{appendix:num_method}.

\section{Setting the stage}
\label{sec:eoms}

In this section, we introduce the equations of motion (EOMs). We then illustrate the initial angular configurations adopted for neutrinos and antineutrinos.

\subsection{Neutrino mean field equations}

For simplicity,  we consider  two flavors of neutrinos and refer the reader to Refs.~\cite{Shalgar:2021wlj,Chakraborty:2019wxe,Capozzi:2022dtr,Capozzi:2020kge} for investigations dedicated to three-flavor effects. The evolution of the neutrino flavor field can be modeled in terms of Wigner transformed $2\times2$ density matrices, $\varrho(\vec{p},t)$ for neutrinos and $ \bar{\varrho}(\vec{p},t)$ for antineutrinos. Since we are interested in exploring FFC, for simplicity we ignore the energy dependence of the density matrices, and hence $\vec{v}=\vec{p}/E$ is a unit vector. Moreover, we impose axial symmetry on the initial configurations and the solutions---see  Ref.~\cite{Shalgar:2021oko} for details on symmetry breaking effects in non-axially symmetric systems. 

The vector $\vec{v}$ is defined with respect to the symmetry axis (zenith angle $\theta$), while the azimuthal angle $\phi$ has been integrated out. The velocity component along  the symmetry  axis is $v=|\vec{v}|\cos{\theta}=\cos{\theta}$, since $|\vec{v}|=1$ for (anti)neutrinos traveling at the speed of light. The velocity component takes values between $v=1$ (forward direction) and $v=-1$ (backward direction). 
 
After these simplifications, the neutrino density matrix is
\begin{eqnarray}
\varrho(v,t) = 
\begin{pmatrix}
\varrho_{ee}(v,t) & \varrho_{ex}(v,t)\\
\varrho_{ex}^{*}(v,t) & \varrho_{xx}(v,t) 
\end{pmatrix}\ ,
\end{eqnarray}
whose diagonal elements represent the occupation numbers of neutrinos of different species, while the off-diagonal terms contain information about  coherence between flavors. An analogous expression holds for the density matrix associated to antineutrinos, $\bar{\varrho}(v,t)$. 
 
\begin{figure*}
\includegraphics[width=0.49\textwidth]{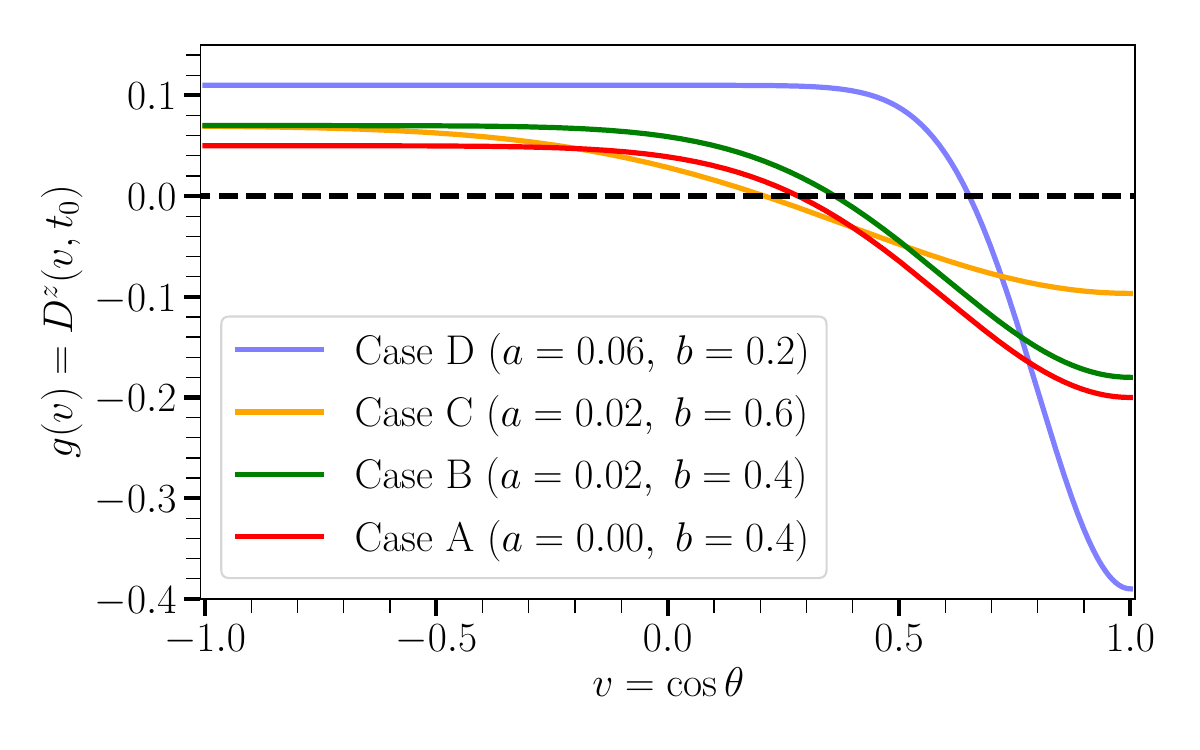}
\includegraphics[width=0.49\textwidth]{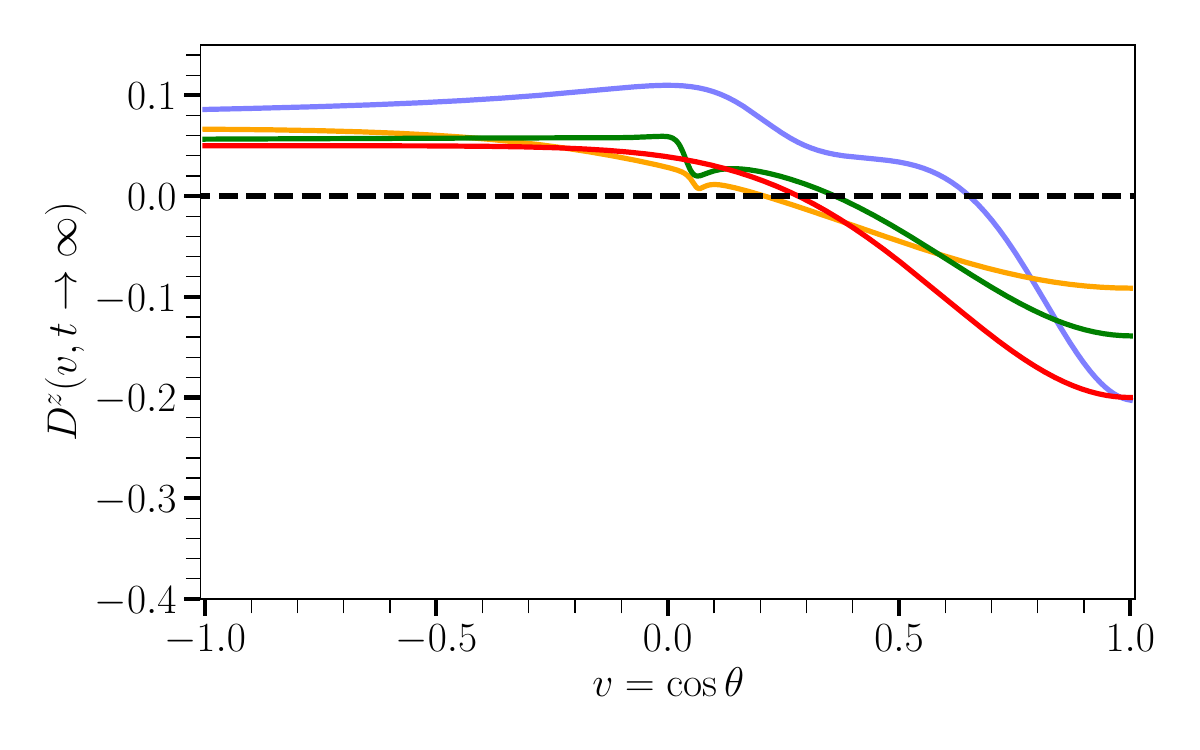}
\caption{\textit{Left:} Representative ELN angular distributions $g(v)$, following Eqs.~\eqref{eq:eln_distr} and \eqref{eq:eln_distr1}. \textit{Right:} ELN angular distributions in the presence of damping at $t=5\times 10^{-6} \ \mathrm{s^{-1}}$, essentially corresponding to the final distribution at $t\to\infty$. The ELN angular distributions $D^z(v,t)$ evolve in such a way that the total lepton number $D^z_0=\int dv D^z(v,t)$ is a constant of motion  with the initial $D^z_0=\int dv\, g(v)$, i.e., the integrals over corresponding curves in the left and right panel are the same. 
}
\label{fig:1}
\end{figure*}
 
The  EOMs for neutrinos and antineutrinos are respectively
\begin{eqnarray}
\label{eq:eoms}
\kern-1em (\partial_t+\vec{v}\cdot \vec{\nabla}) \varrho(v,t) &=&-i [H_{\nu\nu}(v,t),\varrho(v,t)] + \mathcal{C}(\varrho(v,t)),  \\
\kern-1em (\partial_t+\vec{v}\cdot \vec{\nabla})\bar{\varrho}(v,t) &=&-i [\bar{H}_{\nu\nu}(v,t),\bar{\varrho}(v,t)] +  \bar{\mathcal{C}}(\bar{\varrho}(v,t)).
 \label{eq:eoms1}
\end{eqnarray}
The term on the left-hand side of the EOMs is the advective operator, $\vec{v}\cdot \vec{\nabla}$, which affects flavor evolution if the medium is inhomogeneous~\cite{Shalgar:2019qwg,Wu:2021uvt, Sigl:2021tmj,Richers:2021xtf}. In this work, we neglect the advective term and focus on homogeneous cases. 

On the right-hand-side of the EOMs, the neutrino self-interaction Hamiltonian is responsible for the development of flavor transformation:
\begin{eqnarray}
 H_{\nu\nu}(v,t) = \mu \int dv^\prime[\varrho(v^\prime,t)-\bar{\varrho}(v^\prime,t)][1-vv^\prime]\ ;
\end{eqnarray}
it couples neutrinos of different momenta and renders the flavor evolution nonlinear. Since we focus on FFC, we neglect the vacuum and matter terms in the Hamiltonian; thus, the Hamiltonian for neutrinos and antineutrinos is identical i.e.\  $H=\bar{H}=H_{\nu\nu}$. However, we refer the reader to Refs.~\cite{Shalgar:2020xns,Padilla-Gay:2021ywy,Abbar:2017pkh,Johns:2021qby} for work dedicated to explore the impact of these terms on the FFC phenomenology. 

The second term on the right-hand side of the EOMs is the collision term $\mathcal{C}$, which takes non-forward neutrino scattering with the background medium into account. Different implementations of the collision term and its impact on the flavor conversion phenomenology have been explored in the recent literature \cite{Capozzi:2018clo,Johns:2021qby,Shalgar:2022rjj,Shalgar:2022lvv,Hansen:2022xza,Martin:2021xyl,Shalgar:2020wcx,Kato:2022vsu,Sasaki:2021zld,Sigl:2021tmj,Richers:2019grc}. We here use a minimal model where random collisions are assumed to have the effect of damping the flavor coherence of a given momentum mode without adding or removing particles so that~\cite{PhysRevD.36.2273,Raffelt:1992uj}
\begin{eqnarray}\label{eq:iC}
\mathcal{C}(\varrho(v,t)) = -\Gamma 
\begin{pmatrix}
0 & \varrho_{ex}(v,t)\\
\varrho_{ex}^{*}(v,t) & 0
\end{pmatrix}  \ . 
\end{eqnarray}
An analogous expression holds for $\bar{\mathcal{C}} = \alpha {\mathcal{C}}$ where the parameter $\alpha$ accounts for the possibility that quantum damping effects act differently on neutrinos and antineutrinos. 
For the moment we assume equal damping rates ($\alpha=1$), whereas later (Sec.~\ref{sec:diff_damping}), we will also explore the scenario of different damping rates ($\alpha \neq 1$).

\subsection{EOMs for lepton and particle number}

FFC is primarily a phenomenon of the lepton-number distribution and it turns out the EOMs strongly simplify if one considers the density matrices for lepton and particle number instead of the ones for neutrinos and antineutrinos. We therefore define the sum (S, particle number) and difference (D, lepton number) matrices~\cite{Johns:2020qsk,Johns:2021qby,Padilla-Gay:2021haz}
\begin{eqnarray}\label{eq:SvDv}
 \sS(v,t) &=& \varrho(v,t)+\bar{\varrho}(v,t) \ , \\ 
 \sD(v,t) &=& \varrho(v,t)-\bar{\varrho}(v,t) \ .
\end{eqnarray}
A special role is played by the density matrix of total lepton number and the one for lepton-number flux
\begin{eqnarray}\label{eq:D0}
 \sD_0&=&\int_{-1}^{+1} dv\,\sD(v,t)\ ,
 \\ \label{eq:D1}
 \sD_1&=&\int_{-1}^{+1} dv\,v\,\sD(v,t)\ .
\end{eqnarray}
Moreover, we express all $2\times2$ density matrices by Bloch vectors in the usual way. Then the EOMs are
\begin{eqnarray}\label{eq:DS_generala}
\kern-1.2em\dot{\vD}(v) &=& -\mu\vD(v)\! \times\!\vD_0 + \mu v\vD(v)\!\times\!\vD_1 -\Gamma \vD^{xy}(v)\ , \\
\kern-1.2em\dot{\vS}(v) &=& -\mu\vS(v) \times\vD_0 +\mu v\vS(v) \times\vD_1 -\Gamma \vS^{xy}(v)\ , 
    \label{eq:DS_general1a}
\end{eqnarray}
where we define the ``transverse'' vectors as  
$\vD^{xy}(v)=(D^{x}(v), D^{y}(v), 0)$ and $\vS^{xy}(v)=(S^{x}(v), S^{y}(v), 0)$. They represent the off-diagonal terms of the density matrices and thus the flavor coherence of a given mode.

Integrating Eq.~\eqref{eq:DS_generala} over $dv$ immediately reveals that \smash{$\dot{D}^z_0=0$} and \smash{$\dot\vD_0^{xy}=-\Gamma \vD_0^{xy}$} in the presence of damping. If the system is initialized with vanishing transverse components, the total flavor lepton number of the system is conserved.
Therefore, even in the damped case, the first term in both equations only causes  a simple precession around the conserved vector $\vD_0$ without affecting the internal dynamics of the system or the question of flavor conversion. Henceforth we drop this term (we go the co-rotating frame) and use
\begin{eqnarray}\label{eq:DS_general}
\dot{\vD}(v) &=& \mu v\vD(v)\!\times\!\vD_1 -\Gamma \vD^{xy}(v)\ , \\
\dot{\vS}(v) &=& \mu v\vS(v) \times\vD_1 -\Gamma \vS^{xy}(v)\ , 
    \label{eq:DS_general1}
\end{eqnarray}
as our EOMs. Our initial conditions are that all $\vD(v)$ are parallel to the $z$-axis (except for a small seed to trigger the instability) and are given by the ELN spectrum
\begin{equation}
    D^z(v,t_0)=g(v)\ ,
\end{equation}
where 
\begin{equation}
    g(v)=\varrho_{ee}(v,t_0)-\varrho_{xx}(v,t_0)-\bar{\varrho}_{ee}(v,t_0)+\bar{\varrho}_{xx}(v,t_0)\ .
\end{equation}
Note that, in the following, we omit the explicit time dependence for simplicity unless otherwise specified. 

Without damping, in Paper~I \cite{Padilla-Gay:2021haz} we found a formal analogy of the EOMs with the ones of a gyroscopic pendulum. In this scenario, $\vD_{0}$ corresponds to the (constant) gravitational field, which exerts a torque on the pendulum $\vD_{1}$, making it swing away from the flavor axis and convert flavor. With damping, $\vD_0$ is still a constant of  motion, resulting in net ELN  conservation, regardless of the initial configuration. While $\vD_{1}$ still performs pendulum-like oscillations, its length now shrinks, as opposed to the earlier
undamped solutions (see for instance Fig.~2 of Paper~I).

\subsection{System setup}

For fast flavor instabilities to occur,  a sufficient condition is the presence of a crossing in the ELN angular distribution~\cite{Morinaga:2021vmc}. However, the presence of flavor instability does not necessarily imply large flavor conversion~\cite{Padilla-Gay:2020uxa, Padilla-Gay:2021haz,Richers:2021xtf,Bhattacharyya:2020jpj,Martin:2019gxb}. For the sake of simplicity, we restrict ourselves to the case of single-crossed families of ELN distributions where $\varrho_{xx}(v,t_0)=\bar\varrho_{xx}(v,t_0)=0$, whereas the electron-flavored terms are of the form: 
\begin{eqnarray}\label{eq:eln_distr}
\varrho_{ee}(v,t_0) &=& 0.50\ ,   \\
\bar{\varrho}_{ee}(v,t_0)&=&  0.45-a + \frac{0.1}{b}   \exp{\Bigg[\frac{-(1-v)^2}{2 b^2}\Bigg]}\ .
\label{eq:eln_distr1}
\end{eqnarray}
The shape of the distributions is set by the parameters $a \in [-0.04, 0.12]$ and $b \in [0.1, 1]$, while the normalization is \smash{$\int_{-1}^{+1} dv \varrho_{ee}(v,t_0) = 1$}. In Fig.~\ref{fig:1} we show $g(v)$ for four selected examples Cases A--D that are identical with the benchmark configurations adopted in Paper~I \cite{Padilla-Gay:2021haz}. We ensure the convergence of our results by fixing the number of angular bins to $N_v=1000$. See Appendix~\ref{appendix:num_method} for details on the numerical method.

\section{Impact of collisional damping}\label{sec:impact_damping}

We now turn to the consequences of damping on  flavor conversion physics. In particular, we explore the steady state reached by the system as a function of the damping rate and compare the flavor phenomenology to the one obtained for cases without damping. In this section, we assume equal damping rates for neutrinos and antineutrinos ($\alpha=1$).

\subsection{Example}

The rate of FFC is governed by the neutrino-neutrino interaction energy $\mu=\sqrt2\GF n_\nu$, whereas weak interaction rates are parametrically $\GF^2 n_{\rm target} E_\nu^2$ and thus $\Gamma\simeq \mu\,\GF E_\nu^2$. For a reference energy of $E_\nu=100$~MeV, we find $\GF E_\nu^2\simeq10^{-7}$, so we are safely in the regime of weak damping. Of course, if the damping rate were large compared to the instability rate, the system would be frozen in analogy to the quantum Zeno effect \cite{Harris:1980zi}. 

In our numerical examples, unless otherwise stated, we use a fixed reference value for the neutrino-neutrino interaction strength $\mu=10^{5} \ \mathrm{km^{-1}} = 3 \times 10^{10} \ \mathrm{s^{-1}}$ and $\Gamma=\Gamma_0 = 2.5\times 10^{6} \ \mathrm{s^{-1}}\simeq 8.3\times 10^{-5}\mu = 8.3 \ \mathrm{km^{-1}}$, and therefore much smaller than $\mu$. So we are in the weak damping regime, yet the asymptotic final state is reached within a reasonable time of evolution. One of our findings will be that in the weak damping regime, the asymptotic final state does not depend on $\Gamma$, although the time to get there is proportional to $\Gamma^{-1}$.

\begin{table}[ht]
\caption{Parameters for our reference examples, where $\cmin$ is the lowest dip of the undamped flavor pendulum and $f$ the asymptotic flux factor as defined in Eq.~\eqref{eq:F}. Cases B--D are unstable, while Case A is stable.}
\smallskip
    \label{tab:costh}
    \centering
    \begin{tabular*}{\columnwidth}{@{\extracolsep{\fill}}lllll}
    \hline\hline
     Case & $a$ & $b$ & $\cmin$ & $f$ \\
     \hline
     A   & 0.00 & 0.4 &   ---    &    --- \\
     B   & 0.02 & 0.4 & $+0.335$ & $0.581$ \\
     C   & 0.02 & 0.6 & $+0.849$ & $0.904$ \\
     D   & 0.06 & 0.2 & $-0.034$ & $0.348$ \\
     \hline
    \end{tabular*}
    \vskip12pt
\end{table}

\begin{figure}[b!]
\centering
\includegraphics[width=0.9\columnwidth]{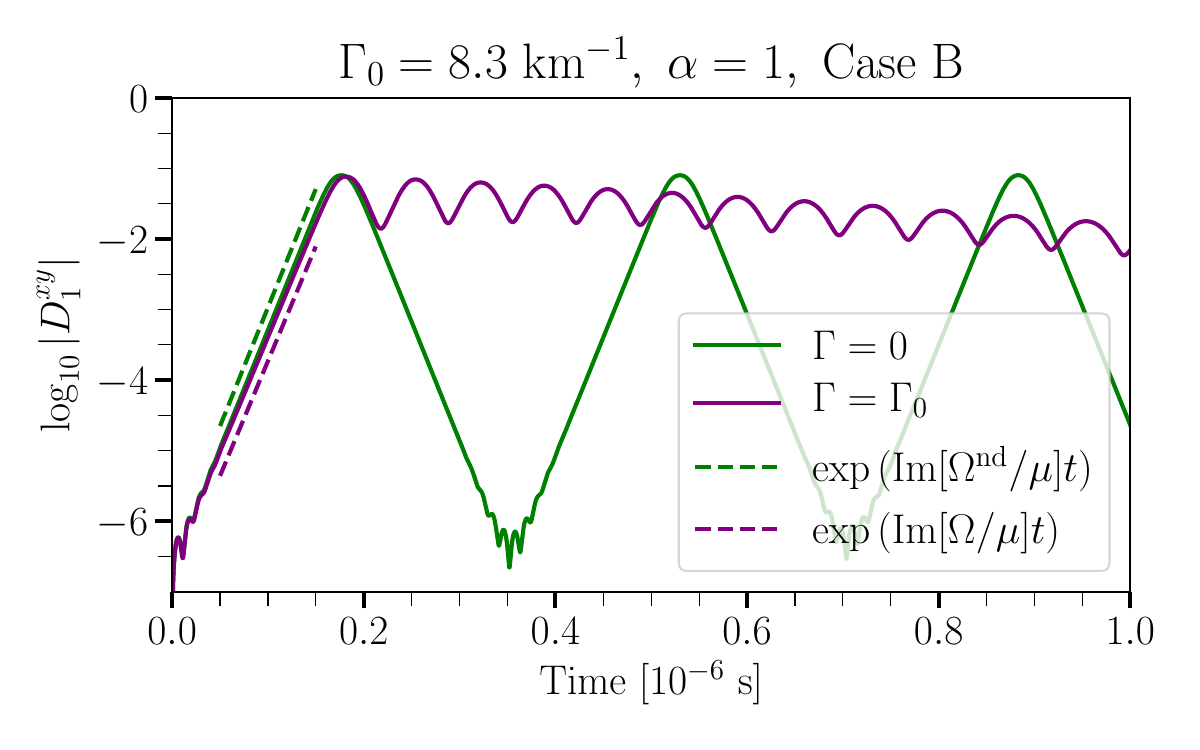}
\includegraphics[width=0.9\columnwidth]{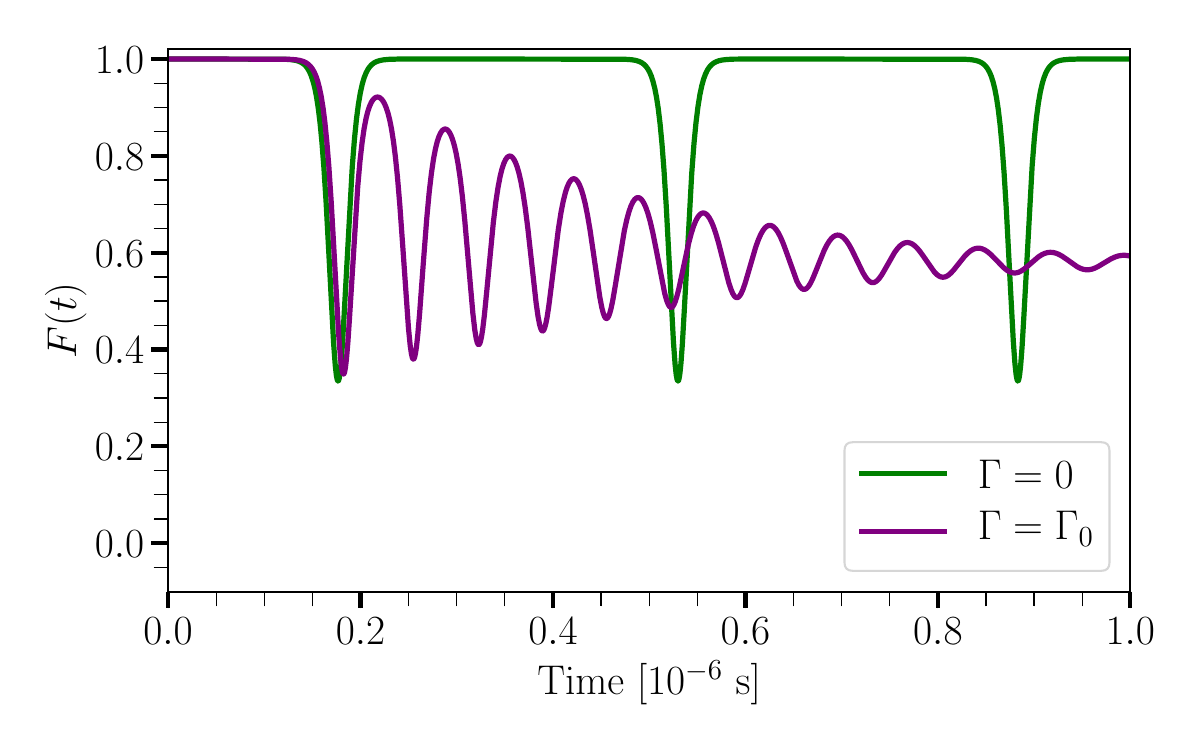}
\includegraphics[width=0.9\columnwidth]{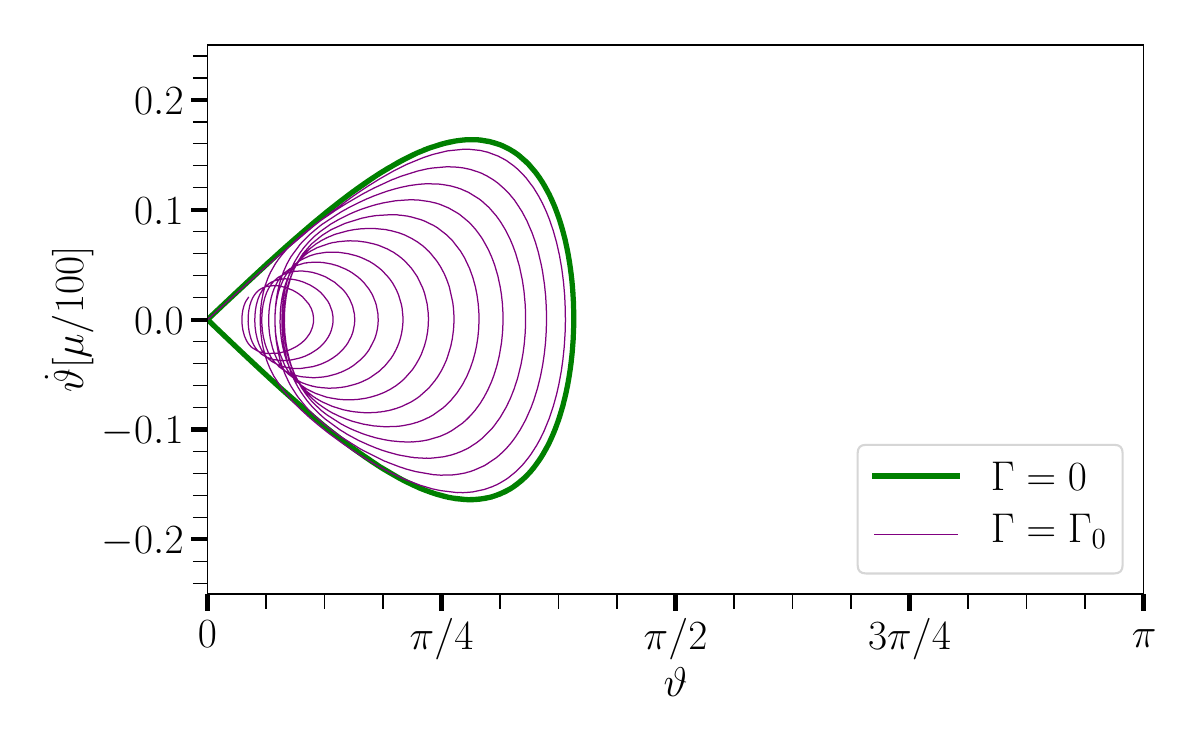}
\caption{Flavor evolution for Case B  without (green) and with (purple) damping, using equal damping rates for $\nu$ and~$\bar{\nu}$. {\it Top panel:} Transverse components $|D_{1}^{xy}|$ (solid) and corresponding growth rates $\mathrm{Im}[\Omega/\mu]$ (dashed).  {\it Middle panel:} ELN flux factor, $F(t)$, where the effect of damping becomes 
visible in the nonlinear regime. While the solution without damping periodically oscillates, the damped solution tends to reach a steady state.  {\it Bottom panel:} Parametric plot of $\dot \vartheta(t)$ as a function  of $\vartheta(t)$. The case without damping encloses the damped solution.
}
\label{fig:3}
\end{figure}

Beginning with Case~B as a first example, we show in the top panel of Fig.~\ref{fig:3} the transverse components $|D^{xy}_{1}|$ (solid lines) and the growth rates (dashed lines) for $\Gamma=0$ (green) and $\Gamma=\Gamma_0$ (purple). The linear regime occurs within the first $2\times 10^{-7}$~s, as visible from the exponential growth of $|D^{xy}_{1}|$. Once the linear phase is over, the solutions with and without damping begin to depart from each other, and damping finally leads to a stationary asymptotic solution after  $\mathcal{O}(10^{-6})$~s, as shown in the middle panel of Fig.~\ref{fig:3}. Without damping, the initial growth rate is $\gamma=\mathrm{Im}[\Omega^{\mathrm{nd}}/\mu]=2.58\times 10^{-3}$. Including damping, it is reduced to $\gamma-\Gamma_0=2.49\times 10^{-3}$, a 4\% reduction for our choice of parameters.

Our benchmark damping rate is small enough that the linear regime is almost unaffected; see the top panel of Fig.~\ref{fig:3}. The latter can also be appreciated by noticing that the first dip in $F(t)$ is almost identical in both cases. After a few iterations (middle panel of Fig.~\ref{fig:3}), the solution with damping saturates and reaches a steady state solution while the case without damping continues to oscillate regularly and indefinitely~\cite{Padilla-Gay:2021haz}. The minimal dip in the undamped case and the final flux factor are given in Table~\ref{tab:costh}, both values being related by Eq.~\eqref{eq:asymptotic}. Indeed, numerically the final steady state does not depend on $\Gamma$ as long as it is small.

In flavor space, one can introduce the angle $\vartheta$ defined as the angle between the flavor axis (also that of $\vD_0$) and the vector $\vD_1$ (see Eq.~8 of Ref.~\cite{Padilla-Gay:2021haz} for more details); thus, $\cos{\vartheta}=1$ corresponds to pure electron-flavor content (and therefore no conversion) while  $\cos{\vartheta}=-1$ corresponds to maximal flavor conversion, as allowed by the total lepton number conservation.
An alternative approach to visualize the flavor evolution for the cases with and without damping is reported in the bottom panel of  Fig.~\ref{fig:3}, where a parametric plot of the angular velocity $\dot{\vartheta}$ is shown as a function of the angular position $\vartheta$. The  solution with $\Gamma=\bar\Gamma=0$ (bottom green line in Fig.~\ref{fig:3}) encloses  the damped solution (purple line) as it reaches the steady state $\dot{\vartheta} \rightarrow 0$, $\vartheta \rightarrow 0$. The   maximum excursion of the pendulum $\vD_1$ from its initial orientation along the flavor axis is given by~\cite{Padilla-Gay:2021haz}:
\begin{eqnarray}\label{eq:costhmin}
 \cmin = -1 + 2\sigma^2 .
\end{eqnarray}
Here, $\sigma$ is the spin parameter that is connected to the spin of the pendulum (no damping) through $S=2\sigma\lambda$, where $\lambda$ is the natural frequency of the pendulum. For a solution to be unstable, the spin parameter $\sigma$ must be such that $\sigma < 1$~\cite{Padilla-Gay:2021haz}. In the example presented in Fig.~\ref{fig:3} (see lower panel), the ELN angular distribution is such that $\sigma=0.817$ therefore  $\cos{\vartheta}_{\mathrm{min}}=0.335$ (or $\vartheta_{\mathrm{min}} \approx \pi/3$). When the damped solution reaches the steady state,  the pendulum $\vD_1$ is oriented along the flavor axis and conversions cease. 

\begin{figure*}
\centering
\includegraphics[width=0.49\textwidth]{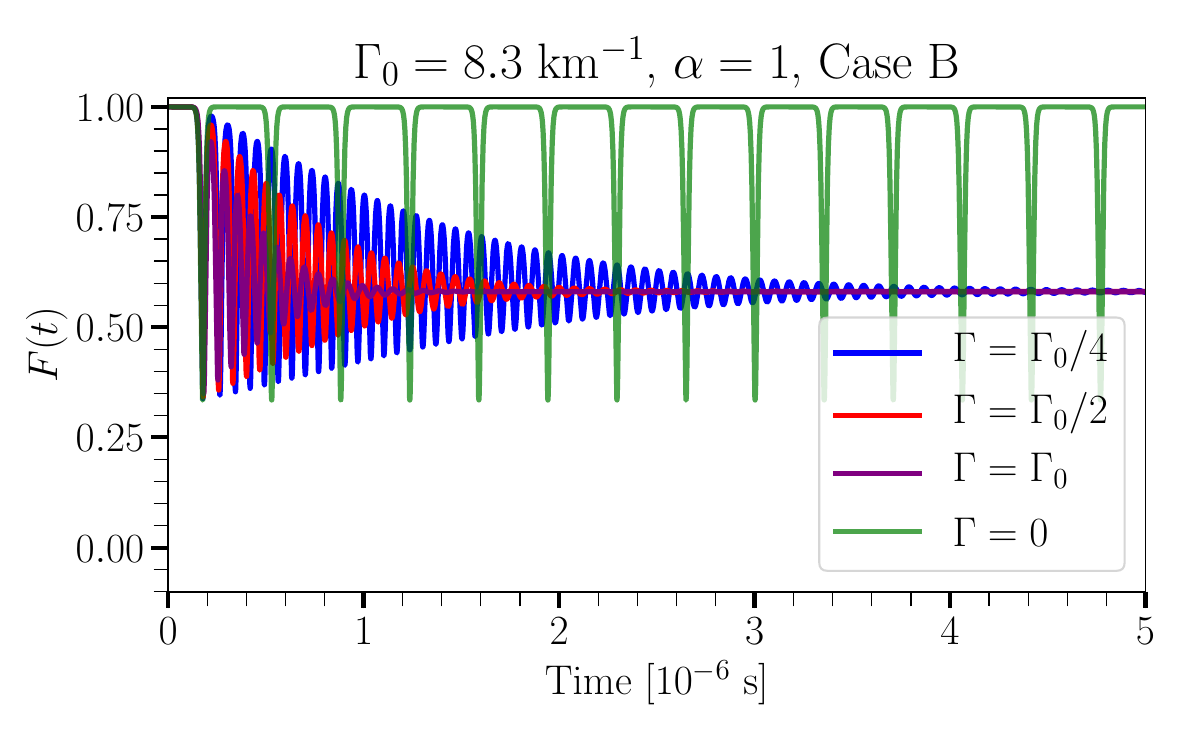}
\includegraphics[width=0.49\textwidth]{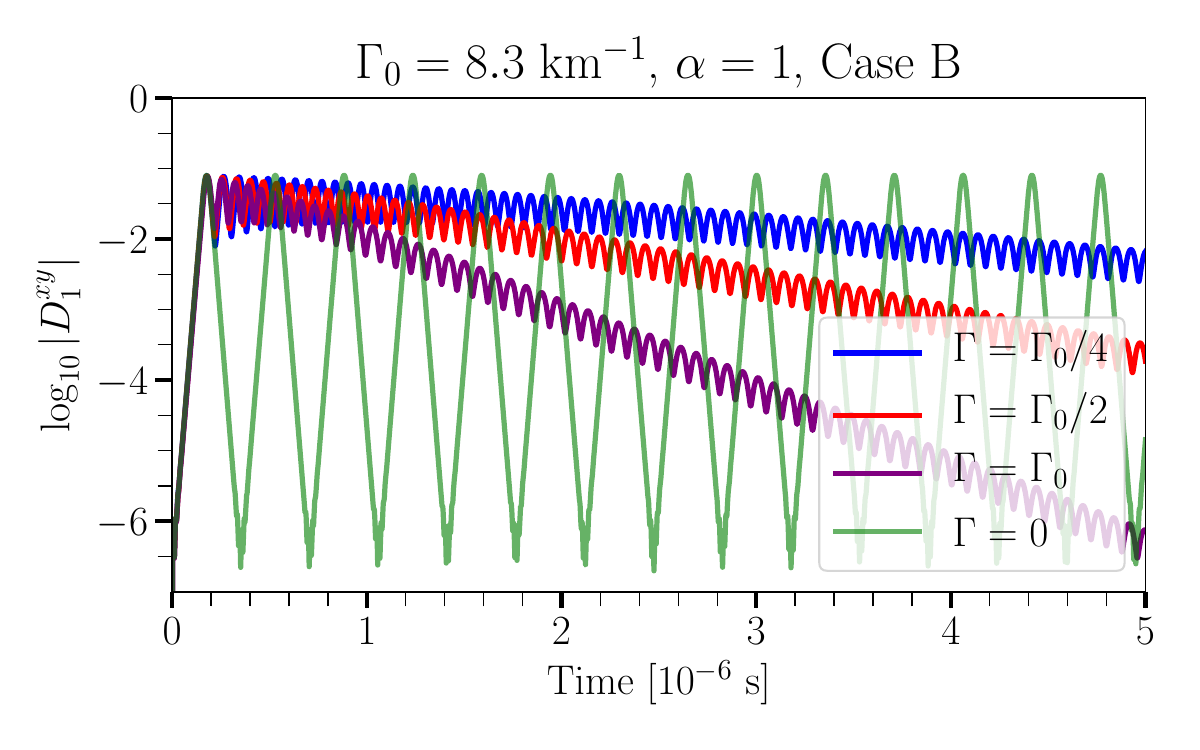}
\includegraphics[width=0.49\textwidth]{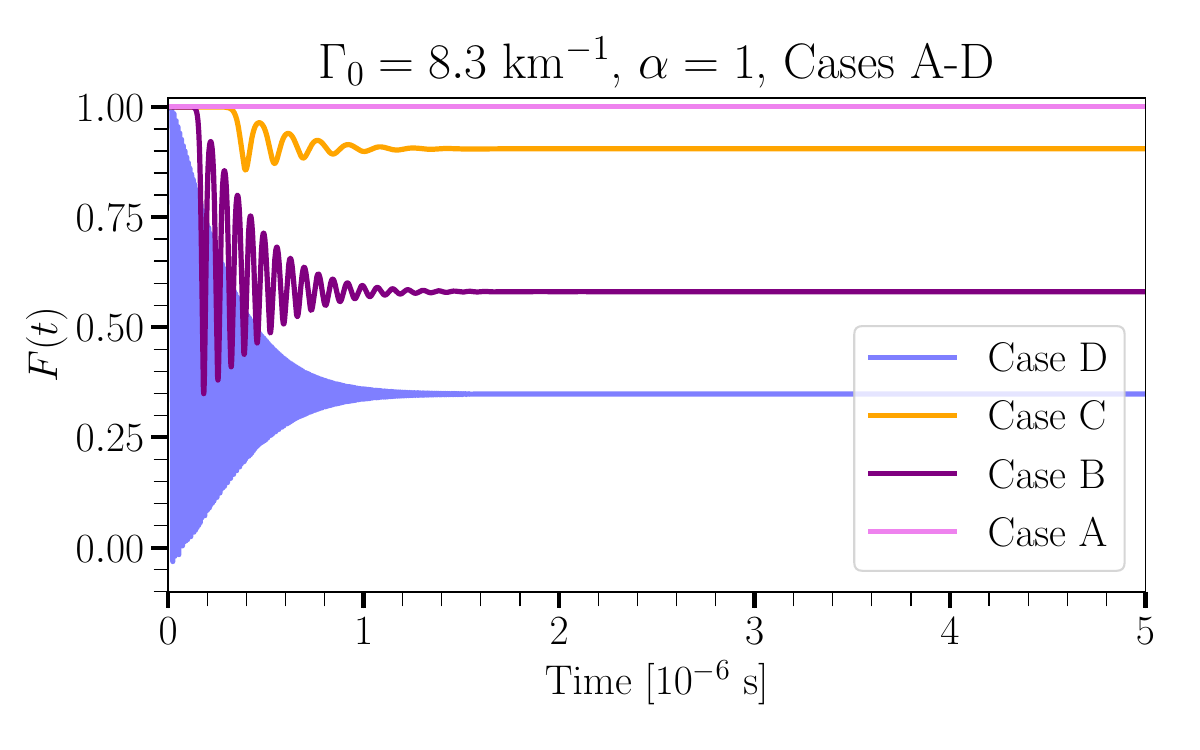}
\includegraphics[width=0.49\textwidth]{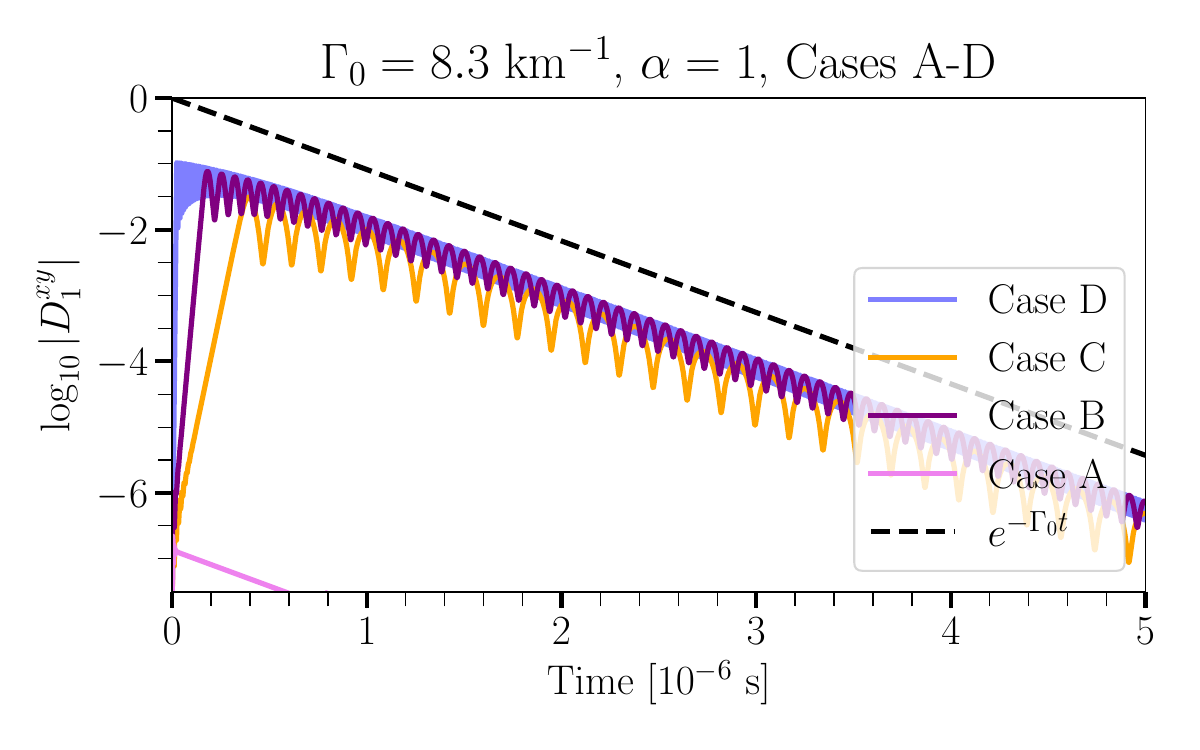}
\caption{{\it Top panels:} Evolution of the ELN flux factor $F(t)$ (left panel) and $D_1^{xy}$ (right panel) for Case B,  same damping rates for neutrinos and antineutrinos, for different values of $\Gamma$. The final flavor outcome is the same,
independently of $\Gamma$, that however determines the time to get there. {\it Bottom panels:} Same for Cases A--D for a single damping rate $\Gamma=\Gamma_0$. The dashed line is $\exp(-\Gamma_0 t)$ and mimics the descent of the transverse component $|D_1^{xy}|$. They all have the same slope, even the stable Case~A, where only the initial seed is damped.}
\label{fig:4a}
\end{figure*}

\subsection{The steady state}

The undamped case is formally equivalent to a gyroscopic pendulum~\cite{Padilla-Gay:2021haz}. For the damped case this remains true in the sense that if we were to stop damping at any time during the evolution, the motion from that time onward is again a gyroscopic pendulum, but one with different parameters (different length $|\vD_1|$, different spin, different natural frequency). The final state then is a gyroscopic pendulum in a stable position.

Mapping the continuous-spectrum case on a pendulum and including damping of the transverse parts of all Bloch vectors, in particular the angular momentum $\vJ$ has a conserved $z$-component which is identical to the spin in a position where the pendulum is oriented along the $z$-axis and thus has no orbital angular momentum. Therefore, the asymptotic final state has the same spin as the initial state, although in between the spin changes and only returns to its initial state towards the end.

Clearly the damped pendulum has conserved parameters that determine the final asymptotic state, but thus far we were not able to find an analytic prediction of the final state properties, which we hence explore numerically. In particular,
the top left panel of Fig.~\ref{fig:4a} shows the final configuration for Case~B and four different values of the damping rate, while the top right panel shows the evolution of the off-diagonal terms. One can see that the time taken from the system to achieve the steady state configuration is fixed by $\Gamma$, without affecting the final flavor outcome. 

The steady-state configuration depends on the initial ELN, as displayed in the bottom panels of Fig.~\ref{fig:4a}. Since the off-diagonal terms can only exponentially shrink, as described by Eq.~\eqref{eq:DS_general}, our collision term cannot lead to a dynamical enhancement of the transverse components $|D_{1}^{xy}|$. In practice, this means that if a configuration is stable in the absence of damping, such as Case A (pink curve in the lower panels of Fig.~\ref{fig:4a}),  damping  ($\bar\Gamma=\Gamma$) will not affect the stability of the solution, as seen by the exponential decrease of the transverse component (pink curve) in the bottom right panel of Fig.~\ref{fig:4a}. However, this is not necessarily the case
when $\bar\Gamma\not=\Gamma$, as we will see in Sec.~\ref{sec:diff_damping}.

Empirically we have found that the final configuration does not depend on the initial seed nor $\Gamma$, but only on the original pendulum parameters, and more specifically only on the maximum dip reached without damping as provided by 
Eqs.~\eqref{eq:cmin} and \eqref{eq:F} (see the black dotted line in Fig.~\ref{fig:dip} corresponding to the steady state value $f$ (Eq.~\eqref{eq:F}) that we predict empirically). In order to highlight the predictive power of Eq.~\eqref{eq:asymptotic}, Table~\ref{tab:costh} shows  the values of $\cmin$ and $f$ for Cases A--D, which are in perfect agreement with the numerical solutions.

We have tested this empirical result not just for the few Cases~A--D, but have performed many tests. In particular, we have considered the explicit pendulum equations given in Paper~I and included damping of the transverse component of the pendulum vector $\vR(t)=\vD_1(t)$ and $\vJ(t)$, varying the seed, the damping rate, and especially the initial spin, i.e., the initial $\vJ^z$ that returns in the end to its initial value.

\section{Different damping rates for neutrinos and antineutrinos}\label{sec:diff_damping}
In this section, we investigate the flavor conversion phenomenology in the context of different damping rates for neutrinos and antineutrinos. We also show that if $\Gamma \neq \bar\Gamma$, collisions can induce a flavor instability even in the absence of flavor mixing, confirming the findings of Ref.~\cite{Johns:2021qby}.

\subsection{Collisional instability}
The equations of motion for the case of unequal damping rates  $\mathcal{\bar{C}}=\alpha\mathcal{C}$ are a set of coupled differential equations for the vectors $\vS(v)$ and $\vD(v)$ with the following structure:
\begin{eqnarray}\label{eq:DS_alpha}
  \dot{\vD}(v) &=&   \mu v\vD(v) \times \vD_1 - \frac{\Gamma}{2} (1+\alpha)\vD^{xy}(v)  \nonumber \\
  & & -\frac{\Gamma}{2} (1-\alpha)\vS^{xy}(v) \ , \\
  \dot{\vS}(v) &=&  \mu v \vS(v) \times \vD_1  - \frac{\Gamma}{2}  (1+\alpha)\vS^{xy}(v) \nonumber \\ 
  & & -\frac{\Gamma}{2} (1-\alpha)\vD^{xy}(v) \ .
  \label{eq:DS_alpha1}
\end{eqnarray}
For  $\alpha = 1$, the equations for $\vD(v) $ are a closed set of equations~\cite{Padilla-Gay:2021haz,Johns:2019izj} as opposed to the case with $\alpha \neq 1$. The fact that $\vD(v) $ and $\vS(v) $ couple to each other for $\alpha \neq 1$ gives rise to flavor instabilities where there were none for $\alpha = 1$. Such a system was also investigated by Ref.~\cite{Johns:2022yqy}, whose Eqs.~(22) coincide with our Eqs.~\eqref{eq:DS_alpha} and~\eqref{eq:DS_alpha1}.

\begin{figure}
\centering
\includegraphics[width=0.51\textwidth]{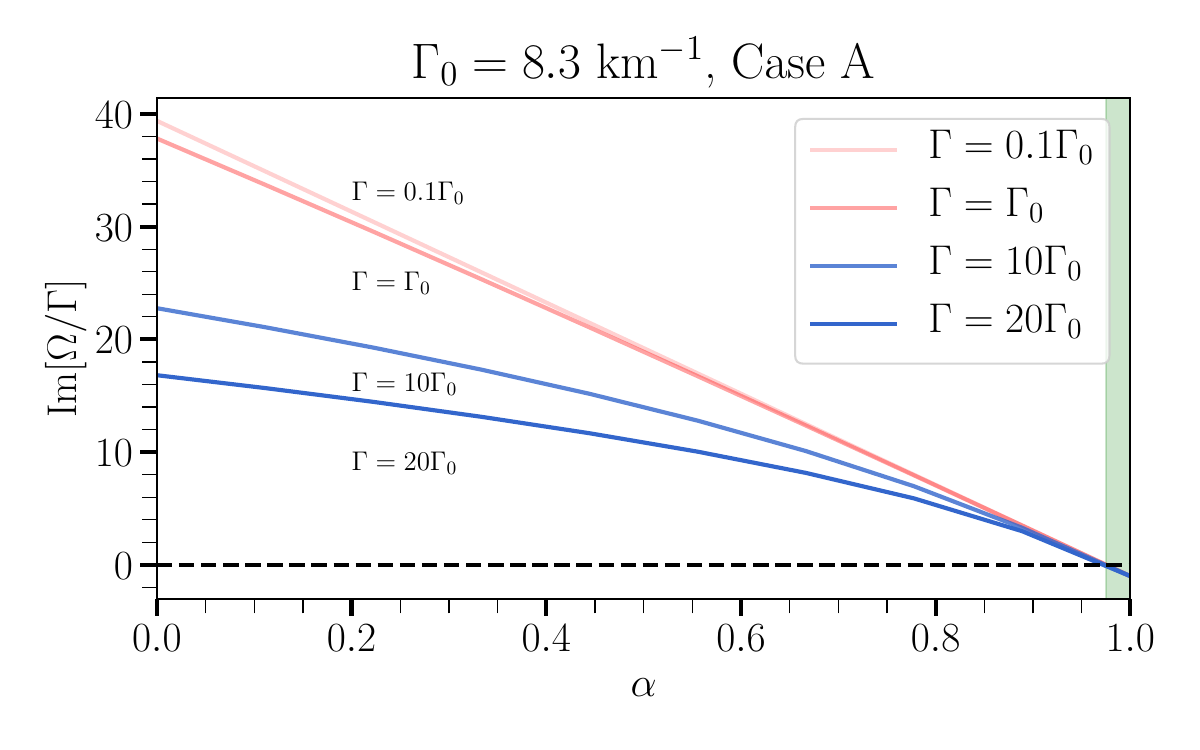}
\includegraphics[width=0.50\textwidth]{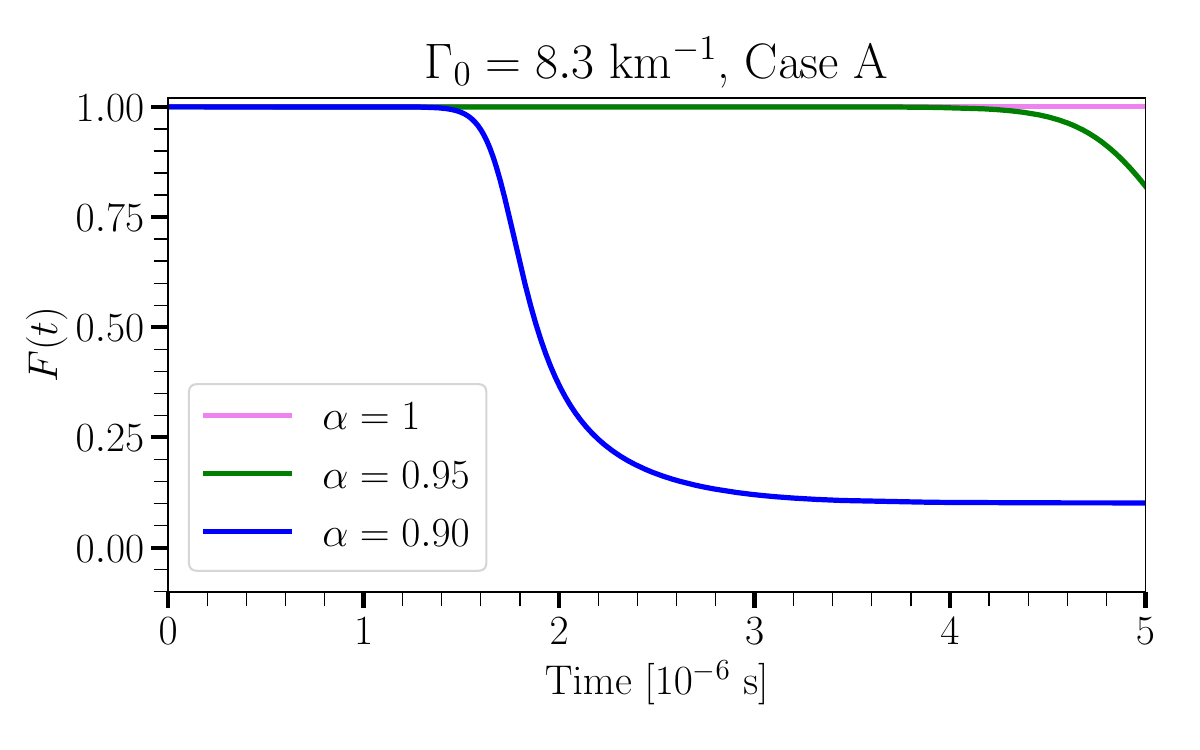}
\caption{Flavor evolution for Case A under the assumption of different damping rates for neutrinos and antineutrinos, i.e. $\alpha\neq 1$. {\it Top panel:} Growth rate as a function of $\alpha$. Case~A is stable in the absence of damping and with damping for $\alpha=1$. For $\alpha=1$, one finds $\mathrm{Im}[\Omega]=-\Gamma$, whereas  the solutions grow exponentially (unstable solutions) for $\alpha < \alpha_{\mathrm{crit}}$, which we found to be $\alpha_{\mathrm{crit}} = 0.975$. The stable region is marked by the green shaded area where $\alpha \geq \alpha_{\mathrm{crit}}$. The value of $\alpha_\mathrm{crit}$ is independent of the chosen value of $\Gamma$ as long as $\Gamma \ll \mu$. {\it Bottom panel:} Temporal evolution of $F(t)$ for three selected values  of $\alpha$. If $\alpha\neq 1$ the flavor solutions become unstable.}
\label{fig:9}
\end{figure}

 We  focus on Case A,  which is stable for $\mathcal{C}=\mathcal{\bar{C}}=0$ and $\mathcal{\bar{C}}=\mathcal{C}$ ($\alpha=1$) and explore its flavor evolution for  $\mathcal{\bar{C}}=\alpha\mathcal{C}$ ($\alpha \neq 1$).
Figure~\ref{fig:9} shows the effect of the coupling between $\vD(v)$ and $\vS(v)$ as visible from  Eqs.~\eqref{eq:DS_alpha} and \eqref{eq:DS_alpha1}. In the top panel of Fig.~\ref{fig:9}, we show the dependence of the growth rate $\mathrm{Im}[\Omega/\Gamma]$ on the magnitude of $\alpha$. For $\alpha=1$, the growth rate equals $\mathrm{Im}[\Omega]=-\Gamma$  and the solution is stable, in agreement with Fig.~\ref{fig:4a}. Eventually, $\alpha$ reaches the critical value $\alpha_{\mathrm{crit}} = 0.975$ for which the growth rate transitions from $\mathrm{Im}[\Omega]<0$ to $\mathrm{Im}[\Omega]>0$, rendering the solution unstable. The growth rates of Case A are larger for smaller values of $\alpha$; hence systems with asymmetric damping rates between neutrinos and antineutrinos lead to solutions that grow faster. Moreover, the upper panel of Fig.~\ref{fig:9} shows that the critical value $\alpha_{\mathrm{crit}}$ only depends on the ELN angular distribution and not on the specific value of the damping rate $\Gamma$.

In the bottom panel of Fig.~\ref{fig:9}, we show the evolution of the ELN flux factor $F(t)$ for three different values of $\alpha$. Even for a modest asymmetry between the neutrino and antineutrino damping rate  ($\alpha=0.95 < \alpha_{\mathrm{crit}}$, green line), a significant amount of flavor is converted in the first $5\times 10^{-6}$~s. A more extreme asymmetry ($\alpha=0.90< \alpha_{\mathrm{crit}}$, blue line) leads to near flavor equipartition within the same time interval. 

\begin{figure}
\centering
\includegraphics[width=0.49\textwidth]{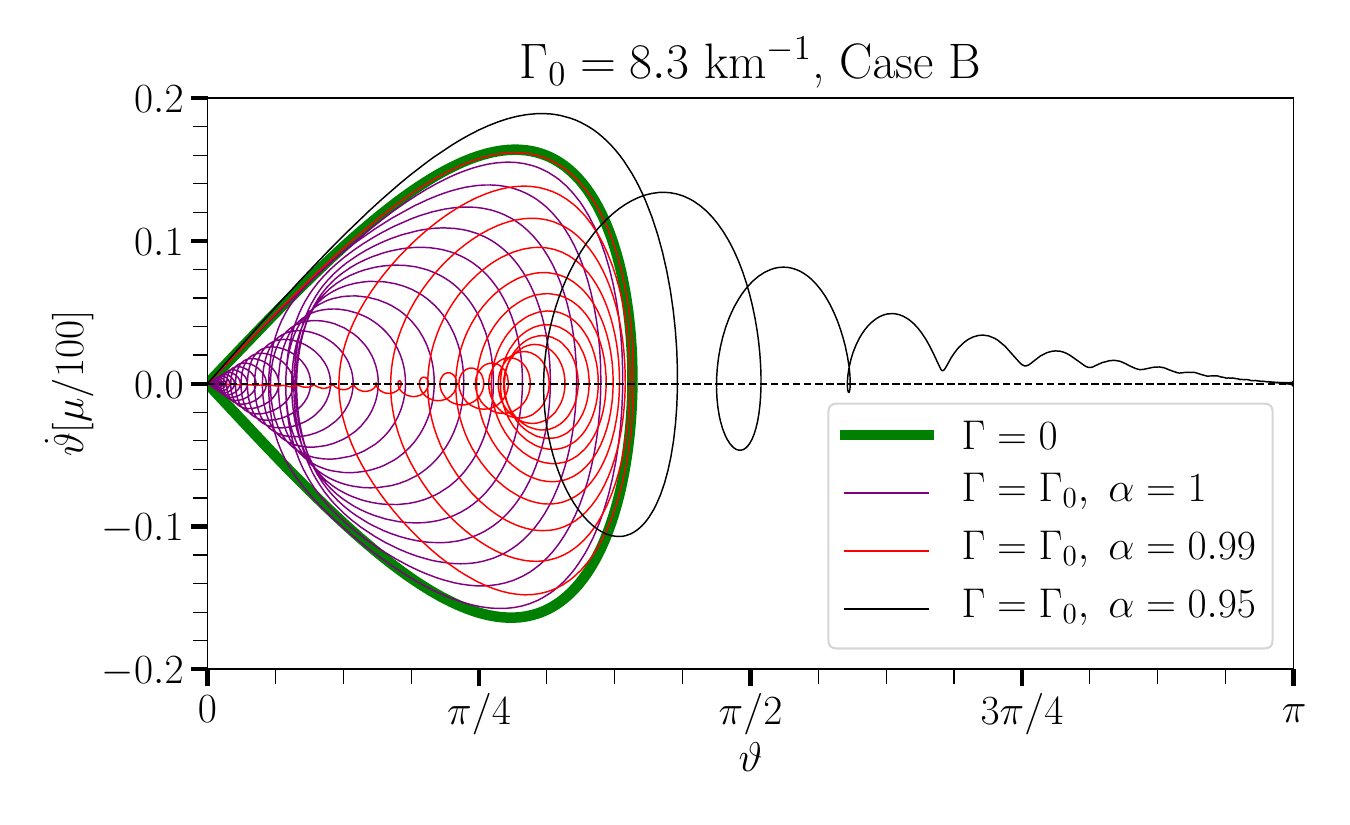}
\caption{Parametric plot of $\dot\vartheta$ as a function of $\vartheta$  for Case B for different damping rates for neutrinos and antineutrinos.  For  $\alpha \neq 1$, the trajectory escapes the ``envelope'' defined by the pendulum solution (green line), flipping the orientation of the polarization vector while shrinking.}
\label{fig:9a}
\end{figure}

In  Fig.~\ref{fig:9a}, we show the phase-space dynamics for different values of $\alpha$ for Case B. For  $\alpha\neq 0$, the phase space trajectories escape the ``envelope'' subtended by the undamped solution (green line). For instance, for  $\alpha=0.95$, the phase-space trajectory escapes the envelope during the first iterations of the evolution, reaching its steady state $(\dot{\vartheta}=0)$. At that point, the polarization vector $\vD_1$ points downwards ($\vartheta=\pi$) in contrast to the $\alpha=1$ case where $\vD_1$ points upwards ($\vartheta=0$).

\subsection{Stability criteria}\label{sec:criteria}

The conditions for flavor instability can be found by studying an isotropic system that consists only of two modes $\vP$ and $\bar{\vP}$ for simplicity. Within this system, flavor instabilities are not triggered by the ELN angular crossing, but rather by the asymmetry between damping rates, as originally pointed out in Refs.~\cite{Johns:2021qby,Johns:2022yqy}. Nevertheless, our results carry over to the ELN-crossed configurations (i.e.~Case A) considered in Fig.~\ref{fig:9}. For an isotropic system the EOMs (Eq.~\eqref{eq:DS_generala}) can be integrated over the velocity variable $v$ to obtain
\begin{eqnarray}
    \dot{\vP} &=& \mu (\vP-\vPb)\times \vP - \Gamma \vP^{xy} \ , \\
    \dot{\vPb} &=& \mu (\vP-\vPb)\times \vPb - \Gamma \alpha \bar{\vP}^{xy} \ , 
\end{eqnarray}
where the ELN density vector is $\vD_0=\int dv (\vP_v-\bar{\vP}_v)=\vP-\bar{\vP}$, whereas
the ELN flux vector vanishes $\vD_1=\int dv v (\vP_v-\bar{\vP}_v) = 0$.  Further simplification of the EOMs leads to 
\begin{eqnarray}
    \dot{\vP} &=& -\mu\vPb\times \vP - \Gamma \vP^{xy} \ , \\
    \dot{\vPb} &=& -\mu \vPb\times\vP  - \Gamma \alpha \bar{\vP}^{xy} \ . 
\end{eqnarray}
The equations for the transverse components are
\begin{eqnarray}
\label{eq:Px}
    \dot{P}^x &=&  \mu (\bar{P}^z P^y - \bar{P}^y P^z) - \Gamma P^x \ , \\
    \dot{P}^y &=&  \mu (\bar{P}^x P^z - \bar{P}^z P^x) - \Gamma P^y \ .
\label{eq:Py}
\end{eqnarray}
Let us define the following linear combinations
\begin{eqnarray}\label{eq:epsilonSD}
    \epsilon &=& P^{x} - i P^{y}\ ,  \\
    \bar{\epsilon} &=& \bar{P}^{x} - i \bar{P}^{y}\ .
    \label{eq:epsilonSD1}
\end{eqnarray}
By combining Eqs.~\eqref{eq:Px} and~\eqref{eq:Py}, one can obtain a set of equations for the transverse components $\epsilon$ and $\bar{\epsilon}$. The linearization assumes that the neutrino transverse component $\epsilon$ is much smaller than $P^{z}$ (and similar for antineutrinos) so that higher order terms such as  $\epsilon^2$  can be dropped. Thus, the equations of motion for the transverse components are
\begin{eqnarray}
   \dot{\epsilon}  &=& - i \mu (P^z \bar{\epsilon}-\bar{P}_z \epsilon) - \Gamma \epsilon \ , \\
   \dot{\bar{\epsilon}}  &=& - i \mu (P^z \bar{\epsilon}-\bar{P}_z \epsilon) - \Gamma \alpha \bar{\epsilon}\ .
\end{eqnarray}
This set of equations can be written in matrix form
\begin{eqnarray}\label{eq:system_two_mode}
\begin{bmatrix}
\dot{\epsilon} \\
\dot{\bar{\epsilon}} 
\end{bmatrix}
=
(-i) M
\begin{bmatrix}
\epsilon \\
\bar{\epsilon}
\end{bmatrix} \ ,
\end{eqnarray}
where the matrix $M$ is given by
\begin{eqnarray}
M = 
\begin{bmatrix}
 -\mu\bar{P}^z -i \Gamma & \mu P^z \\
-\mu \bar{P}^z &  \mu P^z -i\alpha\Gamma
\end{bmatrix}\ .
\end{eqnarray}
The determinant equation $\mathrm{det}(M-\Omega 1_{2\times2}) = 0$ guarantees that there are nontrivial solutions to the linear equations in Eq.~\eqref{eq:system_two_mode}. The determinant equation can be solved analytically, revealing the  eigenfrequency common to $\epsilon$ and $\bar{\epsilon}$, since we have assumed that both neutrinos and antineutrinos evolve collectively i.e.~$\epsilon(t) = \epsilon(t_0)e^{-i\Omega t}$ and~$\bar{\epsilon}(t) = \bar{\epsilon}(t_0)e^{-i\Omega t}$.
\begin{figure}
\centering
\includegraphics[width=0.5\textwidth]{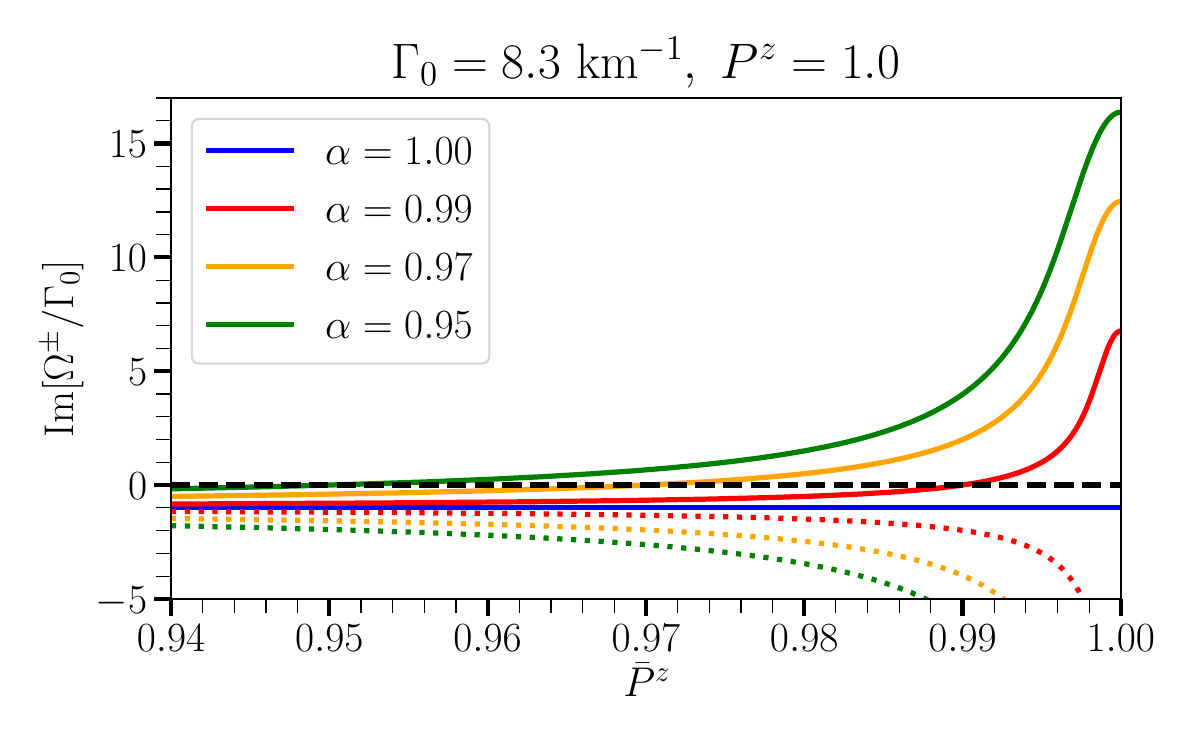}
\caption{Imaginary components of the positive and negative eigenfrequencies $\Omega^{\pm}$ (Eq.~\eqref{eq:omegapm}) for the (isotropic) two mode system. We show $\mathrm{Im}[\Omega^{\pm}/\Gamma_0]$ as a function of $\bar{P}^z$ for different values of $\alpha$ while keeping the other parameters fixed. The components of $\Omega^+$ are shown as solid lines, while the ones for $\Omega^-$ are plotted as dotted lines.  For $\alpha=1$ (blue) the system is stable and $\mathrm{Im}[\Omega^{\pm}/\Gamma_0]<0$ always. On the contrary, for $\alpha\neq 1$ the growth rates reach a maximum when $\bar{P}^z\rightarrow P^z = 1$, independently of $\alpha$ or $\Gamma$ as suggested by the linear expressions for $\mathrm{Im}[\Omega^{\pm}]$ in Eq.~\eqref{eq:IMomegapm_lin}. The instability transition occurs at $\bar{P}^z = \alpha_{\mathrm{crit}}$, making solutions unstable for the interval $\bar{P}^z > \alpha_{\mathrm{crit}}$ in all configurations.}
\label{fig:omegapm}
\end{figure}
The determinant equation is a quadratic equation in the eigenfrequency $\Omega$ with the following pair of solutions:
\begin{eqnarray}\label{eq:omegapm}
\Omega^{\pm} &=&   \frac{\mu}{2}D^z -\frac{i \Gamma}{2}(1+\alpha) \nonumber \\ 
& & \pm \frac{1}{2} \sqrt{(D^z)^2 \mu^2 -(1-\alpha)^2 \Gamma^2 + 2i S^z(1-\alpha)\Gamma \mu } \ , \nonumber \\
& &
\end{eqnarray}
with the usual definitions for the sum and the difference of the polarization vectors  $S^z=P^z+\bar{P}^z$ and $D^z=P^z-\bar{P}^z$, respectively. 

These solutions $\Omega^{\pm}$ reveal that one needs non-vanishing $\mu \Gamma$ to have solutions that grow exponentially~\cite{Johns:2021qby}; if either $\mu$ or $\Gamma$ are zero, the imaginary components of the eigenfrequencies are either $\mathrm{Im}[\Omega^{\pm}]<0$ or $\mathrm{Im}[\Omega^{\pm}]=0$, respectively. Moreover, one can clearly see that for the case with equal damping rates for neutrinos and antineutrinos (i.e.~$\alpha=1$) the solutions $\Omega^{\pm}$ reduce to $\Omega^{-} = - i \Gamma$ and $\Omega^{+} = \mu D^z - i \Gamma $ where $\mathrm{Im}[\Omega^{\pm}]=-\Gamma$, in agreement with our results in Fig.~\ref{fig:9}.

We can approximate the expression for the roots $\Omega^{\pm}$ for the limiting case of weak damping, i.e.~$\Gamma \ll \mu$. By Taylor expanding the square root in Eq.~\eqref{eq:omegapm} we obtain the following expression for the eigenfrequencies:
\begin{eqnarray}\label{eq:omegapm_lin}
\Omega^{\pm} &=& \frac{\mu}{2}D^z -\frac{i \Gamma}{2} \frac{(1+\alpha)D^z \pm (1-\alpha)S^z}{D^z} + \mathcal{O}( {\Gamma^2}/{\mu^2}) \ , \nonumber \\
& &
\end{eqnarray}
where, without loss of generality, we can assume that $\bar{P}^z<P^z$ and $\alpha \in [0,1]$ to ensure that all quantities are positive. Neglecting higher order terms $\sim \mathcal{O}( {\Gamma^2}/{\mu^2})$ and rearranging terms lead to
\begin{eqnarray}\label{eq:IMomegapm_lin}
\mathrm{Im}[\Omega^{-}] &=&  - \Gamma \frac{P^z-\alpha \bar{P}^z}{P^z-\bar{P}^z} \ , \nonumber \\
\mathrm{Im}[\Omega^{+}] &=&  + \Gamma \frac{\bar{P}^{z}-\alpha P^z }{P^z- \bar{P}^z} \ .
\end{eqnarray}
 The solution that grows exponentially is $\Omega^{+}$. Thus, the stability transition occurs when $\mathrm{Im}[\Omega^+]=0$ which occurs if and only if
\begin{eqnarray}\label{eq:alphacrit}
\alpha_{\mathrm{crit}}  = \frac{\bar{P}^{z}}{P^z} \ .
\end{eqnarray}

Hence, solutions are unstable for $\alpha < \alpha_{\mathrm{crit}}$. Notice that, in the small $\Gamma$ limit, the dependence of $\mathrm{Im}[\Omega^{\pm}]$ on $\alpha$ is linear, which explains the behavior shown in the upper panel of Fig.~\ref{fig:9}. Furthermore, the parameter $\alpha$ in Eq.~\eqref{eq:alphacrit} controls the slope of $\mathrm{Im}[\Omega^\pm]$ while the relative ratio between $\bar{P}^z$ and $P^z$ determines where the growth rate crosses zero.

One can reproduce the eigenfrequencies shown in the upper panel of Fig.~\ref{fig:9} using such a two mode system with $P^z=1$ and $\bar{P}^z=0.975$; such choice implies $\alpha_\mathrm{crit}=0.975$ according to Eq.~\eqref{eq:alphacrit}. Finally, in light of the expressions for $\Omega^{\pm}$ (Eq.~\eqref{eq:omegapm}) we find that the growth rates are maximal when $\bar{P}^z \rightarrow P^z$, regardless of the particular choice of $\alpha$ and $\Gamma$, as shown in Fig.~\ref{fig:omegapm}. Notice that all solid lines in Fig.~\ref{fig:omegapm} cross zero at $\bar{P}^z = \alpha_{\mathrm{crit}}$ indicating that they follow the instability condition given by Eq.~\eqref{eq:alphacrit} which in this case is simply $\bar{P}^z > \alpha_{\mathrm{crit}}$, as shown in Fig.~\ref{fig:omegapm}.

\section{Dependence of the final flavor configuration from the ELN crossing}\label{sec:dependence_ELN}

\begin{figure}
\centering
\includegraphics[width=0.45\textwidth]{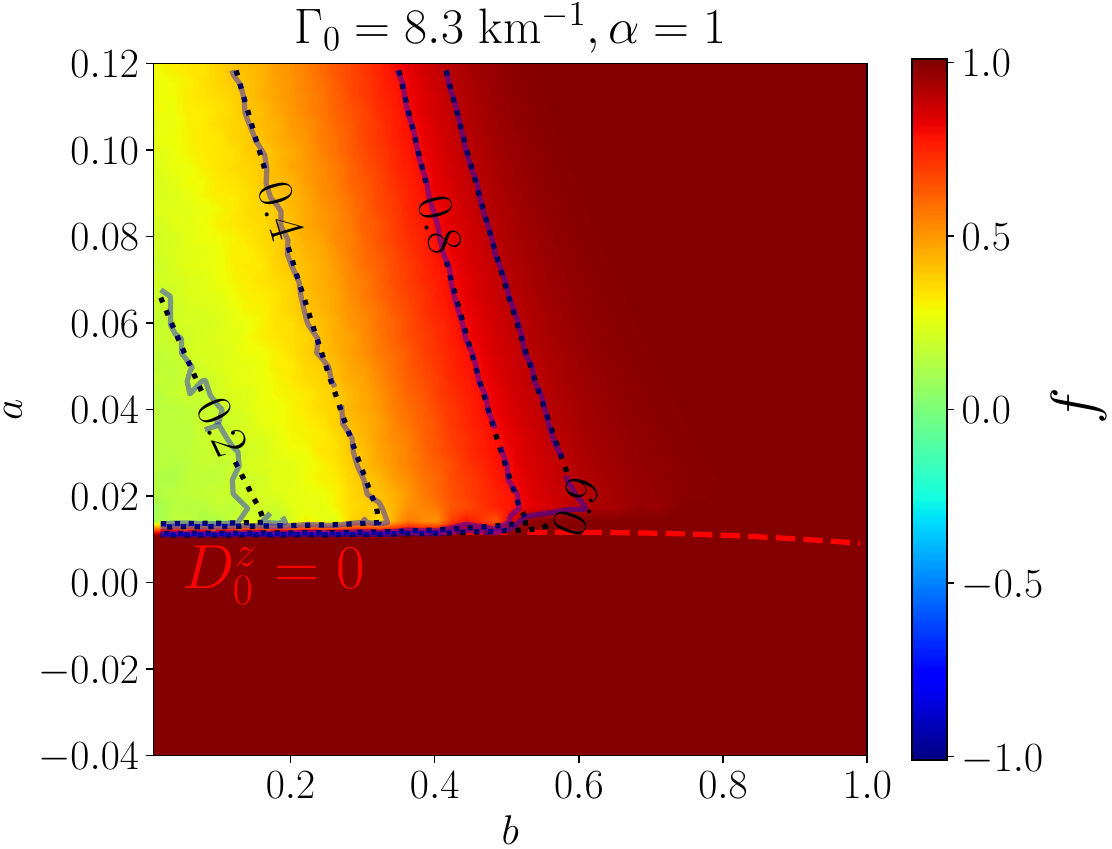}
\includegraphics[width=0.45\textwidth]{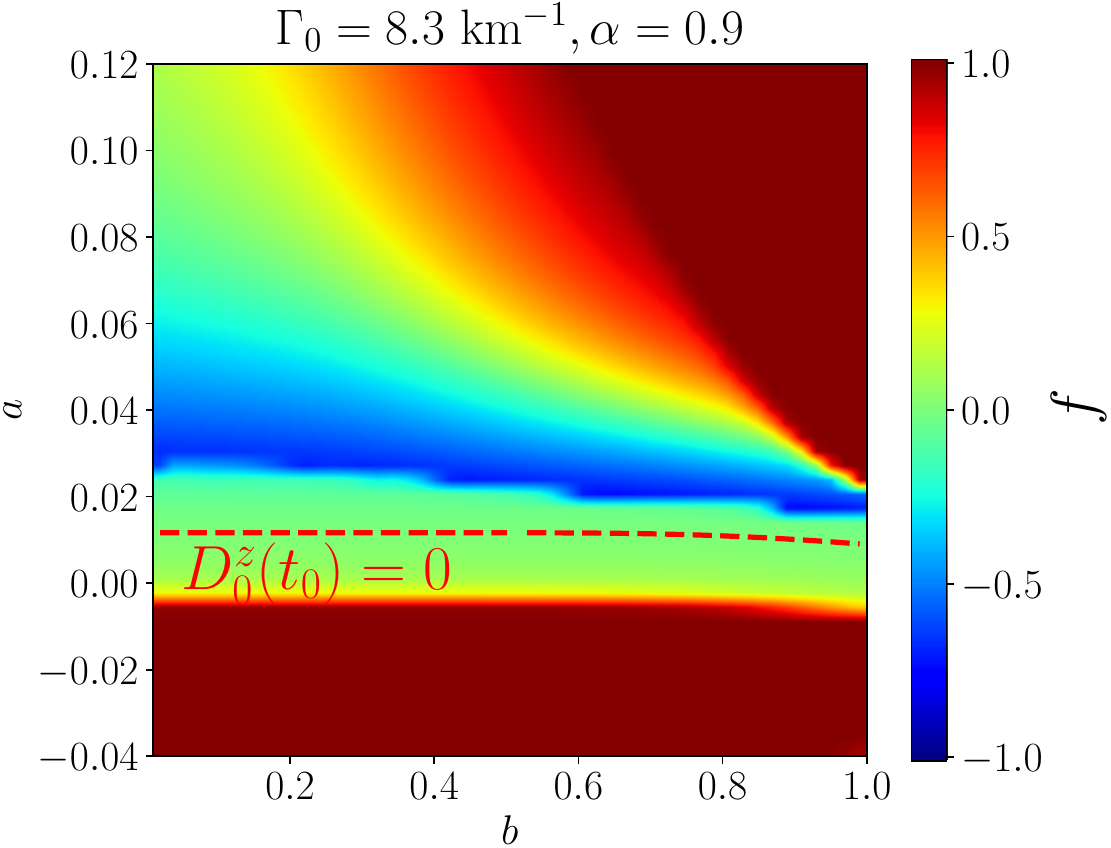}
\caption{Flavor evolution for a wide range of single-crossed ELN spectra characterized by the parameters $a$ and $b$ (see Eqs.~\eqref{eq:eln_distr} and \eqref{eq:eln_distr1}). In the top panel we present the case of same damping rates for neutrinos and antineutrinos $(\Gamma = \Gamma_0, \alpha=1)$, whereas in the bottom panel we show the scenario with unequal damping rates for neutrinos and antineutrinos $(\Gamma = \Gamma_0 ,\alpha=0.9)$. The locus of vanishing lepton number $D_{0}^{z}=0$ is marked with a red dashed line. In the colormaps, we show the steady state value $f$ (Eq.~\eqref{eq:asymptotic}) after $2.5\times 10^{-6}$~s. As shown in the middle panel, the contours of the predicted value $f$ (dotted line) and the numerical solution (solid line) are in excellent agreement for a wide range of single-crossed ELN distributions. The red regions show little to no conversions, while blue shows significant flavor transitions. In agreement with Fig.~\ref{fig:9}, the $(\Gamma = \Gamma_0, \alpha = 0.9)$ case displays the largest regions of the parameter space where flavor conversion occurs, including cases otherwise stable when  $\Gamma= 0$ or $\alpha = 1$.
\label{fig:14}}
\end{figure}

In this section, we generalize our findings to a family of ELN distributions, modeled as described in Eqs.~\eqref{eq:eln_distr} and \eqref{eq:eln_distr1}. The ensemble of  ELN angular configurations determined by the parameters $a$ and $b$ can be represented with a two-dimensional grid as shown in Fig.~\ref{fig:14}. The parameter $a$ controls the relative normalization between neutrinos and antineutrinos, while $b$ parametrizes how forward-peaked the angular distributions are. Running over a range of values of $a$ and $b$ allows us to explore a family of single-crossed ELN spectra systematically. Each point in the plane represents an ELN configuration for which we compute $f$ through Eq.~\eqref{eq:asymptotic}. We find that the analytical results match very well the final flavor outcome obtained numerically. 

Figure~\ref{fig:14} shows contour plots of the asymptotic value $f$ computed in the plane spanned by the parameters $a$ and $b$. The simulation is stopped after $\Delta t = 2.5 \times 10^{-6} \ \mathrm{s}$ to ensure that all configurations have reached a steady state as confirmed by Fig.~\ref{fig:4a}. We performed resolution tests to ensure that the reached steady state is a reliable result; for more details on the numerical methods see Appendix~\ref{appendix:num_method}. We consider the following two scenarios: equal damping rates for neutrinos and antineutrinos ($\alpha=1$, top panel), and unequal damping rates for neutrinos and antineutrinos ($\alpha=0.9$, bottom panel). Although not shown here, for $\Gamma=0$ a large portion of the parameter space has an average value $\langle F(t)\rangle $ close to $1$ since conversions are periodic and no asymptotic value is ever reached, as discussed in Paper I.

The parameter space is divided by the $D_0^z=P^z-\bar{P}^z=0$ contour (red dashed line) which separates stable solutions from unstable ones~\cite{Padilla-Gay:2021haz}. For the case with equal damping rates for neutrinos and antineutrinos (upper panel of Fig.~\ref{fig:14}), flavor decoherence forces the pendulum to settle around the steady state value $f$ (Eq.~\eqref{eq:asymptotic}), without returning to the $\cos{\vartheta}=1$ (stable) configuration, in agreement with the results reported in Figs.~\ref{fig:3} and~\ref{fig:4a}. Qualitatively, $f$ reaches $f\simeq 0$ for a large region of the parameter space, especially for forward-peaked ELN configurations (small values of $b$). In agreement with the stability criteria in Eq.~\eqref{eq:alphacrit}, we find that the $D_0^z=P^z-\bar{P}^z=0$ transition boundary remains unchanged since the imaginary component of the growing solution $\Omega^{+}$ (Eq.~\eqref{eq:IMomegapm_lin}) changes sign when $\bar{P}^z=P^z$. We find that the computed value of $f$ is in excellent agreement with our numerical simulations, as shown by the overlap between the black dotted line (our estimation) and the solid blue line (numerical) for a wide range of ELN angular distributions.

The most extreme scenario is the one obtained for unequal damping rates for neutrinos and antineutrinos (bottom panel of Fig.~\ref{fig:14}), where the steady state value of $f$ can even reach negative values (blue region) $f\simeq -0.9$, in agreement with our findings in Fig.~\ref{fig:9} where the vector $\vD_1$ can change its orientation for smaller values of $\alpha$.
For the case with unequal damping rates for neutrinos and antineutrinos, the lepton number is not a constant of motion; see Eq.~\eqref{eq:DS_alpha}. For these systems, we show the locus of initially-vanishing lepton number $D_{0}^{z}(t_0)=0$, which helps visualize the deviation from the pendulum-like solution and well as where new instabilities arise as a result of $\alpha\neq 1$. In the region below $D_{0}^{z}(t_0)=0$, the component $D_{0}^{z}$ is negative and the growing solution is $\Omega^{-}$ (Eq.~\eqref{eq:IMomegapm_lin}). Thus, the stability transition occurs when $P^z-\alpha \bar{P}^z=P^z-0.9\bar{P}^z=0$, which lies barely outside the simulation box and allows for the bottom part of the parameter space to become unstable due to unequal damping rates. In the other cases with equal damping rates (upper panel) or no damping, the lepton number $D_{0}^{z}$ is strictly conserved, and the locus of $D_{0}^{z}=0$ remains constant. No conversions are allowed in this region, in agreement with the stability criteria for the gyroscopic pendulum~\cite{Padilla-Gay:2021haz}.

\section{Conclusions}\label{sec:conclusions}

Our earlier work~\cite{Padilla-Gay:2021haz} shows that it is possible to exploit a formal analogy of the neutrino EOMs with the ones of a gyroscopic pendulum, confirming previous findings reported in Refs.~\cite{Hannestad:2006nj,Duan:2006an,Fogli:2007bk,Johns:2019izj}, allowing us to predict the final flavor configuration analytically for a homogeneous and axially symmetric system. In this work, we follow up on our previous findings and investigate the role of damping due to random collisions in the final flavor configuration. We assume spatial homogeneity and axial symmetry and work in the two-flavor framework. Even in the presence of damping, the ELN lepton number vector, $\vD_0$, is conserved and plays the role of ``gravity,'' exerting a torque on the dynamical ELN flux vector, $\vD_1$. A limitation of our work is the assumption of spatial homogeneity, which enable us to estimate the final flavor outcome. However, an estimation that simultaneously includes space and time evolution is out of the scope of this work. Moreover, the simple form of our collision term that mimics the loss of coherence among neutrino flavors remains an approximation and a more realistic collision term should be considered.

Although no simple gyroscopic pendulum analogy can be found in the presence of damping, most of the features of the gyroscopic pendulum outlined in Ref.~\cite{Padilla-Gay:2021haz} are preserved. We provide a simple analytical formula to estimate the final steady state achieved by the system and show that it is a simple linear function of the one found to predict the lowest point of the pendulum in Ref.~\cite{Padilla-Gay:2021haz}. Our estimations are in excellent agreement with our numerical computations for a wide range of 
single-crossed ELN spectra. 

Under the assumption of equal damping rates for neutrinos and antineutrinos,  the final flavor outcome differs from the scenario without damping. In particular, we find that the same steady state flavor configuration is reached in the presence of damping and independently of the particular values of the damping rate $\Gamma$ (as long as $\mu \gg \Gamma$). However, the time the system takes to reach such a configuration is a function of the damping rate. 

When the damping rates for neutrinos and antineutrinos are different, new regions of flavor instability appear, and systems that were stable in the absence of damping or for equal damping rates for neutrinos and antineutrinos may become unstable. In particular, even in the absence of ELN crossings, we find that unequal damping rates for neutrinos and antineutrinos could trigger flavor instabilities, confirming the findings of Ref.~\cite{Johns:2021qby}. 
The criteria for flavor instabilities triggered by the damping rate asymmetry are provided in this work using an isotropic system, however our findings carry over to anisotropic systems with nontrivial momentum dependence. 

This work provides new insights into the flavor evolution of neutrinos in dense neutrino environments. Our analytical findings shed light on the rich phenomenology of FFC in the presence of random collisions, offering  simple estimations on the final flavor outcome.

\acknowledgments
We would like to thank Sajad Abbar, Rasmus S.L. Hansen, Lucas Johns, and Shashank Shalgar for useful discussions. This project has received support from the Villum Foundation (Project No.~13164), the Danmarks Frie Forskningsfonds (Project No.~8049-00038B), and the Deutsche Forschungsgemeinschaft through Sonderforschungbereich SFB 1258 ``Neutrinos and Dark Matter in Astro- and Particle Physics'' (NDM) and under Germany’s Excellence Strategy through the Cluster of Excellence ORIGINS EXC-2094-390783311.

\begin{appendix}

\section{Normal mode analysis in the presence of damping}
\label{appendix:lsa_continuous}

In this Appendix, we carry out the normal mode analysis in the presence of collisional damping. We note that a complementary study on the stability criteria with collisional damping is provided in Ref.~\cite{Johns:2021qby}, however, one important difference is that our normal mode analysis includes non-isotropic ELN distributions, while the analysis in Ref.~\cite{Johns:2021qby} focuses on isotropic systems. As such, the analysis of this Appendix captures both instabilities due to angular crossings and collisional damping.

We start by linearizing the EOMs and tracking the evolution of the off-diagonal terms:
\begin{eqnarray}\label{eq:ansatz}
 \varrho_{ex}(v) = Q(v)e^{-i\Omega t} \ \mathrm{and}\ 
 \bar{\varrho}_{ex}(v) = \bar{Q}(v)e^{-i\Omega t} ,
\end{eqnarray}

\begin{figure}[b]
\includegraphics[width=0.47\textwidth]{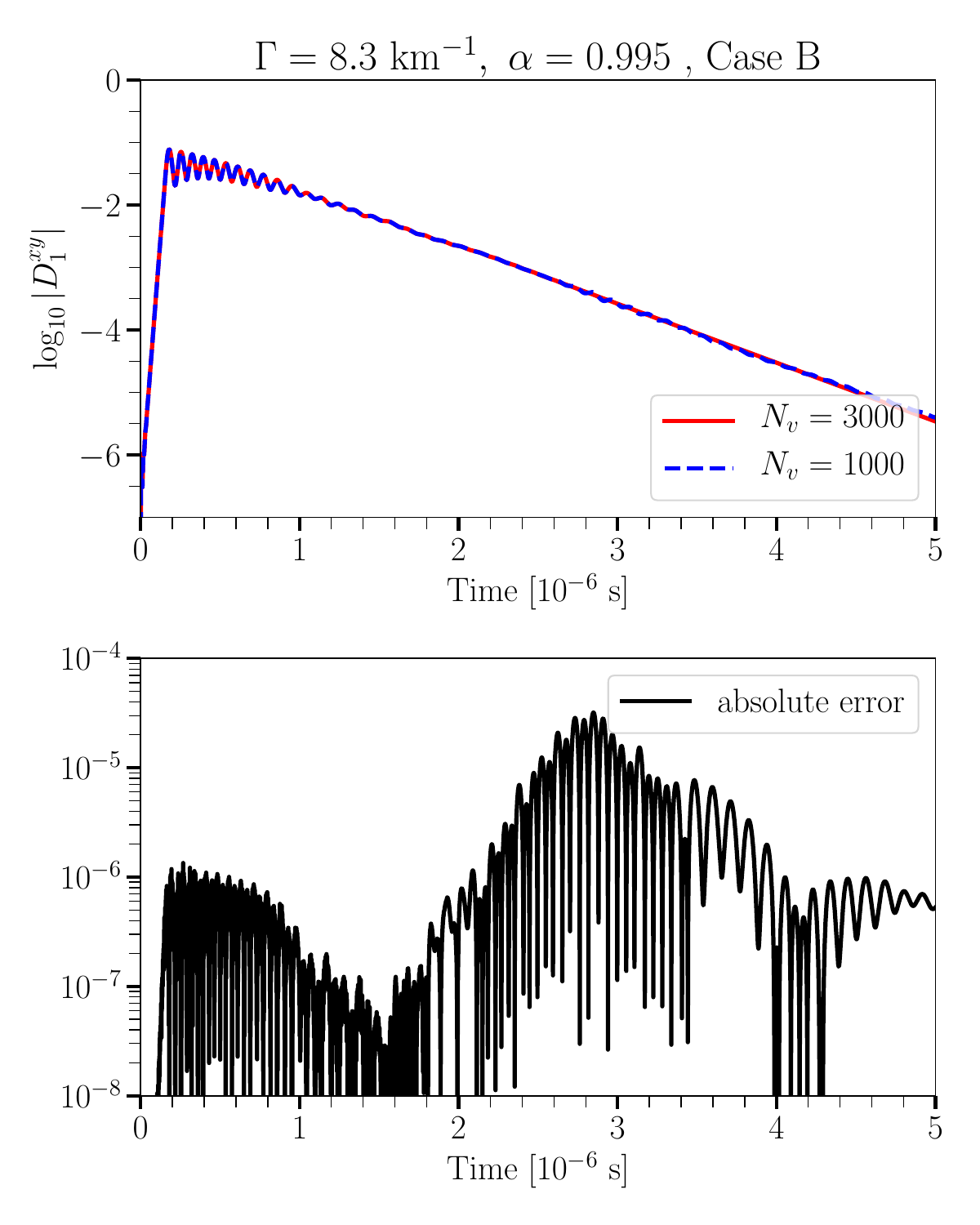}
\caption{\textit{Top panel: } Time evolution of the transverse components $D_1^{xy}$ of Case B with 3000 (solid red) and 1000 angular bins (dashed blue). Here, we focus on the scenario with different damping rates $(\alpha \neq 1)$ as this is the most challenging one from the numerical point of view. \textit{Bottom panel: } Absolute value of the difference between the red and blue solutions of the top panel. The absolute error is about $10^{-5}$ and well under control within the integration time window.}
\label{fig:16}
\end{figure}

where $\Omega$ represents the collective oscillation frequency for neutrinos and antineutrinos. We look for temporal instabilities for the homogeneous mode ($\vec{k}=0$). The off-diagonal component of the EOM for neutrinos and antineutrinos are
\begin{eqnarray}\label{eom_lin}
 i \dot{\varrho}_{ex}(v) &=&  H_{ee}(v)\varrho_{ex}(v) - \varrho_{ee}(v)H_{ex}(v) \nonumber \\ 
 & & -i \Gamma \varrho_{ex}(v) \ , \\
 i \dot{\bar{\varrho}}_{ex}(v) &=&  H_{ee}(v)\bar{\varrho}_{ex}(v) - \bar{\varrho}_{ee}(v)H_{ex}(v) \nonumber \\
 & & -i \alpha \Gamma \bar{\varrho}_{ex}(v) \ ,
\end{eqnarray}
where we have assumed $\varrho_{xx} (t_0) = \bar\varrho_{xx}(t_0) = 0$. Again, here $\Gamma$ is the damping rate, and $\alpha$ allows for a difference in the damping rates of neutrinos and antineutrinos. By substituting Eq.~\eqref{eq:ansatz} in the equation above and solving for $Q(v)$, we obtain 
\begin{eqnarray}\label{eq:Qtheta}
 Q(v) = \frac{ \varrho_{ee}(v) \int d v^\prime [ Q(v^{\prime})-\bar{Q}(v^{\prime}) ] \left[1 - v v^{\prime} \right] } { -\Omega - i \Gamma + A(v) } \ ,
\end{eqnarray}
where we express $\Omega$ and $\Gamma$ in units of $\mu$. Also, we have defined the angle-dependent quantity $A(v)$ as
\begin{eqnarray}
	A(v) & \equiv &  \int d v^\prime [ \varrho_{ee}(v^{\prime})-\bar{\varrho}_{ee}(v^{\prime})] [1- v v^\prime]  \ . 
\end{eqnarray}
A similar procedure follows for $\bar{Q}(v)$: 
\begin{eqnarray}\label{eq:Qthetabar}
 \bar{Q}(v) = \frac{ \bar{\varrho}_{ee}(v) \int d v^\prime [ Q(v^{\prime})-\bar{Q}(v^{\prime}) ] \left[1 - v v^{\prime} \right] } { -\Omega - i \alpha  \Gamma + A(v) } \ .
\end{eqnarray}

Combining the expressions for $Q(v)$ and $\bar{Q}(v)$, we have 
\begin{widetext}
\begin{eqnarray}\label{eq:QmQbar1}
 Q(v)-\bar{Q}(v) = \int d v^\prime  \left[ \frac{\varrho_{ee}(v)}{-\Omega - i \Gamma + A(v)} - \frac{\bar{\varrho}_{ee}(v)}{-\Omega - i \alpha \Gamma + A(v)} \right] [ Q(v^{\prime})-\bar{Q}(v^{\prime}) ] \left[1 - v v^{\prime} \right]\ .
\end{eqnarray}
From the equation above, it must be true that 
\begin{eqnarray}\label{eq:QmQbar2}
 Q(v)-\bar{Q}(v) &=& \left[ \frac{\varrho_{ee}(v)}{-\Omega - i \Gamma + A(v)} - \frac{\bar{\varrho}_{ee}(v)}{-\Omega - i \alpha \Gamma + A(v)} \right] (\beta_1 - \beta_2 v)\ ,
\end{eqnarray}
\end{widetext}
where $\beta_1$ and $\beta_2$ are unknown coefficients. Substituting Eq.~\eqref{eq:QmQbar2} in Eq.~\eqref{eq:QmQbar1}, we obtain a system of equations for the coefficients $\beta_1$ and $\beta_2$: 
\begin{eqnarray}\label{eq:system}
\begin{bmatrix}
\beta_1 \\
\beta_2
\end{bmatrix}
=
\begin{bmatrix}
\mathcal{I}[1] & - \mathcal{I}[v]  \\
\mathcal{I}[v] & - \mathcal{I}[v^2]  
\end{bmatrix}
\begin{bmatrix}
\beta_1 \\
\beta_2 
\end{bmatrix}
=
\mathrm{M}
\begin{bmatrix}
\beta_1 \\
\beta_2 
\end{bmatrix} \ ,
\end{eqnarray}
where the functional $\mathcal{I}[f]$ is
\begin{eqnarray}
\nonumber
 \mathcal{I}[f] = \int d v \left[ \frac{\varrho_{ee}(v)}{-\Omega - i \Gamma + A(v)} - \frac{\bar{\varrho}_{ee}(v)}{-\Omega - i \alpha \Gamma + A(v)} \right] f(v)\ . \\
\end{eqnarray}

The system of equations has a not trivial solution if and only if the following condition is met
\begin{eqnarray}\label{eqn:det}
 \mathrm{det}( \mathrm{M} - 1_{2\times2} ) = 0\ . 
\end{eqnarray}
To search for instabilities, we need to look for the solutions with $\mathrm{Im}[\Omega] \neq 0$. 

If $\alpha=1$, the functional $\mathcal{I}[f] $ simplifies to
\begin{eqnarray}
 \mathcal{I}[f] = \int d v \left[ \frac{\varrho_{ee}(v)-\bar{\varrho}_{ee}(v)}{-\Omega - i \Gamma + A(v)} \right] f(v)\ .
\end{eqnarray}

\section{Numerical methods and convergence}
\label{appendix:num_method}

We implement different initial conditions while keeping the architecture of the numerical simulations unchanged. The angular variable $v$ is discretized in $N_v=1000$ bins. We compute the flavor evolution according to the EOMs (Eqs.~\ref{eq:DS_generala} and~\ref{eq:DS_general1a}). 

The temporal evolution of the system is computed by using an adaptive method. In particular we implement a Runge-Kutta-Fehlberg(7,8) method from the  Boost library~\cite{BoostLibrary}. 

Figure~\ref{fig:16} shows the flavor evolution of a system with damping using distinct angular binning. Both solutions agree very well with each other even in the most challenging scenario of different damping rates $(\alpha \neq 1)$.

\end{appendix}


\bibliography{references.bib}
\end{document}